\documentclass[lettersize,journal, review]{IEEEtran}
\usepackage{times}
\usepackage{epsfig}
\usepackage{graphicx}
\usepackage{amsmath}
\usepackage{amsfonts}
\usepackage{amssymb}
\usepackage{bm}
\usepackage{cite}
\usepackage{multirow}
\usepackage{caption}
\usepackage[utf8]{inputenc}
\usepackage{graphicx}
\usepackage{subcaption}
\usepackage{adjustbox}

\graphicspath{{Figures/}}

\usepackage[pagebackref=true,breaklinks=true,letterpaper=true,colorlinks,bookmarks=false]{hyperref}

\begin{document}

\title{Approximately Invertible Neural Network for Learned Image Compression}

\author{Yanbo Gao, Meng Fu, Shuai Li\textsuperscript{*},~\IEEEmembership{Senior Member,~IEEE}, Chong Lv, Xun Cai, \\Hui Yuan,~\IEEEmembership{Senior Member,~IEEE}, Mao Ye
\thanks{Manuscript received Jun. 19, 2024.}
\thanks{This work was supported in part by the National Natural Science Foundation of China under Grant 62271290, 62001092 and 61901083; and in part by SDU QILU Young Scholars Program. \textsuperscript{*}Corresponding Author: Shuai Li}
\thanks{Yanbo Gao, Meng Fu, Shuai Li, Chong Lv, Xun Cai, Hui Yuan are with Shandong University, Jinan, China. E-mail: \{ybgao, shuaili, caixunzh, huiyuan\}@sdu.edu.cn \par Mao Ye is with University of Electronic Science and Technology of China, Sichuan, China. E-mail: cvlab.uestc@gmail.com
 }
}


\maketitle

\begin{abstract}
In recent years, with the development of deep learning, learned image compression have attracted considerable interests. It typically comprises an analysis transform, a synthesis transform, quantization and a probabilistic model for entropy coding. The analysis transform and synthesis transform, which can be regarded as coupled transforms, are used to encode an image to latent feature and decode the quantized feature to reconstruct the image. However, the analysis transform and synthesis transform are designed independently in the existing methods, making them unreliable in high-quality image compression. Inspired by the success of invertible neural networks in generative modeling, invertible modules are used to construct the coupled analysis and synthesis transforms. Considering the noise introduced in the feature quantization invalidates the invertible process, this paper proposes an Approximately Invertible Neural Network (A-INN) framework for learned image compression. It formulates the rate-distortion optimization in lossy image compression when using INN with quantization, which differentiates from using INN for generative modelling. Generally speaking, A-INN can be used as the theoretical foundation for any INN based lossy compression method. Based on this formulation, A-INN with a progressive denoising module (PDM) is developed to effectively reduce the quantization noise in the decoding. Moreover, a Cascaded Feature Recovery Module (CFRM) is designed to learn high-dimensional feature recovery from low-dimensional ones to further reduce the noise in feature channel compression. In addition, a Frequency-enhanced Decomposition and Synthesis Module (FDSM) is developed by explicitly enhancing the high-frequency components in an image to address the loss of high-frequency information inherent in neural network based image compression, thereby enhancing the reconstructed image quality. Extensive experiments demonstrate that the proposed A-INN framework outperforms the existing learned image compression methods and traditional image compression methods.
\end{abstract}


\section{Introduction}
Image compression is a fundamental problem in the field of image processing and communication, with the primary goal of minimizing the redundancy present in image data \cite{MaZJZWW20,10478821,9694511,LeiL0J0G22,9646488,10121344}. Especially for lossy image compression, it is to utilize fewer bits for encoding, while still ensuring the recovery of a high-quality image that closely approximates the original one. Typically, lossy image compression models are composed of an analysis and synthesis transformation pair, a quantizer, and an entropy coder, all designed to optimize the compression process \cite{wallace1991jpeg,bross2021overview,9205606,9359473,DuanLMHMZ24,ChenM23,9455349,9568930}. Traditional compression schemes such as JPEG \cite{wallace1991jpeg}, JPEG2000 \cite{christopoulos2000jpeg2000}, HEVC \cite{sullivan2012overview}, and VVC \cite{bross2021overview} are designed based on handcrafted algorithms, with optimized individual components of the compression process. However, they often lack a holistic approach to joint optimization across all modules, which leads to suboptimal performance. In recent years, the advancements in deep learning and computer vision \cite{xie2021enhanced, JinLPLLH22,JinLPPLL23, 9585549,9934922,9984191,9043584,9998500} have paved the way for end-to-end learned image compression methods. These methods leverage deep neural networks to achieve image compression. Unlike traditional approaches, deep learning-based models facilitate joint optimization of the entire compression pipeline, which has proven to yield higher compression efficiency and surpass the traditional methods \cite{10226331, zou2022devil,9858899,ZhangZT23}.

The current learned image compression methods mostly take the variational autoencoder (VAE) architecture \cite{kingma2013auto,9989403,9144534,10268865}. The VAE model employs an encoder that transforms the raw image data into a low-dimensional continuous-valued latent space feature. This feature is then quantized and encoded to produce a bitstream through an entropy model. During the decoding phase, the compressed bitstream is reconstructed through a reverse process, where the bitstream is first decoded to the feature by the entropy model and then used to reconstruct the image. This end-to-end framework allows for a joint optimization of the different components and the ability of VAE in capturing the underlying data distribution enables it to realize efficient compression.

While VAE-based models have demonstrated impressive performance in image compression, the encoding and decoding are irreversible with unrelated encoder and decoder except the joint training. With images deviated from the data distribution of the training dataset in the practical use, the compression model may not provide good performance. To enhance the reversibility of the compression process, a promising approach is to employ invertible neural networks (INNs) as the core components for the encoder and decoder in image compression as in \cite{helminger2020lossy}. INNs offer a unique advantage by ensuring that the encoding and decoding processes are reversible, allowing for the exact reconstruction of the original image from the latent feature (without considering the quantization). This property is crucial for maintaining high reconstruction quality in compression. However, the invertible design also brings the significant increase of feature channels at the encoder in order to reduce the spatial resolution. Moreover, the effect of quantization process is not considered in the existing INN-based image compression methods. The noise on the encoded feature introduced in the quantization damages the reverse decoding process, thus lowering the reconstruction quality.

To address the above problems, in this paper, we propose a novel end-to-end lossy image compression framework, called Approximately Invertible Neural Network (A-INN). It formulates the rate-distortion optimization in learned image compress with INN and quantization. Accordingly, quantization noise is reduced with a progressive denoising module (PDM), enhancing the overall compression performance. Furthermore, a Cascaded Feature Recovery Module (CFRM) is also developed to learn high-dimensional feature representations from low-dimensional latent features, in order to address the oversmoothing issue associated with feature copying in the INN-based image compression models.

The contributions of this work can be summarized as follows:

\begin{itemize}
\item We present an Approximately Invertible Neural Network (A-INN) for end-to-end lossy image compression. The rate-distortion optimization in learned image compression, using invertible neural network and considering the quantization error, is thoroughly formulated. The difference between A-INN and INN is investigated and demonstrates the importance of A-INN to address noisy features in the framework of invertible flow. The proposed A-INN can be used as the theoretical foundation and plug-and-play module for all INN based lossy compression methods.
  
\item We propose a progressive denoising module (PDM) on top of the invertible flow, where the quantization noise is gradually reduced as in the formulation of A-INN. A Cascaded Feature Recovery Module (CFRM) is also developed to learn the reconstruction of high-dimensional features, in order to further reduce the noise introduced in the feature channel reduction of encoding.

\item We develop a Frequency-enhanced Decomposition and Synthesis Module (FDSM), to mitigate the loss of high-frequency information in image compression. A dual-frequency attention based synthesis is used to enhance the reconstruction by adaptively fusing the low- frequency and high-frequency information.
\end{itemize}

Comprehensive experiments, conducted on the Kodak, Tecnick, CLIC, and CLIC Professional Test datasets, demonstrate that the proposed A-INN framework outperforms the INN baseline and existing learned image compression methods. Ablation study has also been performed, which validates the effectiveness of each proposed module.

\section{Related Work}

\subsection{Lossy Image Compression}
Traditional lossy image compression methods, such as JPEG \cite{wallace1991jpeg}, JPEG2000 \cite{christopoulos2000jpeg2000}, WebP \cite{google2010webp}, BPG \cite{bellard2015bpg}, AV1 \cite{chen2018overview}, HEVC \cite{sullivan2012overview}, and VVC \cite{bross2021overview}, typically consist of several key components including transform, quantization, and entropy coding. For block-based image compression methods such as HEVC and VVC, intra prediction is also used to perform block-based prediction within the image, and the resulted residual is then encoded. The typical process is to use a discrete cosine transform (DCT) \cite{ahmed1974discrete} or discrete wavelet transform (DWT) \cite{marpe2003context} to create a latent representation in the transform domain. This representation is then quantized, and the output discrete values are encoded using a lossless arithmetic coder \cite{netravali1980picture}.The components are usually designed in a handcrafted way and independently from each other, limiting the overall compression performance. 

With the rapid development of deep learning, learned image compression methods based on end-to-end optimization is being actively investigated. Ballé et al. \cite{balle2016end} first presented a learned image compression method using a variational Autoencoder architecture, where the probability distribution of the quantized latent feature was estimated employing a factorized prior model. By leveraging the power of deep learning, the compression result outperforms the traditional compression standard JPEG and JPEG2000. In \cite{balle2018variational}, Ballé et al. further proposed a Gaussian Scale Mixture (GSM) model that utilizes a hyperprior to adaptively formulate the probability distribution of the latent features as a zero-mean Gaussian with estimated variance for entropy coding, capturing the image spatial redundancy. Based on the hyperprior model \cite{balle2018variational}, Minnen et al. \cite{minnen2018joint} extended the GSM model into a conditional Gaussian model, where an autoregressive \cite{van2016conditional} module was developed to help predict the probability distribution based on contextual information. Cheng et al. \cite{cheng2020learned} further proposed to integrate attention modules to process the latent features and replace the original single Gaussian model \cite{minnen2018joint} with discretized Gaussian mixture likelihoods. On top of the above methods, many entropy models have also been proposed, including context-adaptive \cite{lee2018context, zhou2019multi}, channel-wise autoregressive \cite{minnen2020channel}, and parallel \cite{he2021checkerboard} entropy models, which are not further detailed. These developments in entropy modeling contribute to a better estimation of the feature distribution, leading to better compression performance.

In addition to the entropy models, many methods explore different neural network architectures, such as residual networks \cite{li2020deep}, recurrent neural network (RNN) \cite{toderici2017full, johnston2018improved}, Generative Adversarial Network (GAN) \cite{rippel2017real}, and transformers \cite{zhu2021transformer, zou2022devil}, to enhance the Autoencoder. Among these methods, Invertible Neural Network (INN) based compression has also been investigated, which is further described in the following. These architectures aim to enhance the network to learn and represent complex image features, which in turn leads to more effective compression.

\subsection{Invertible Neural Network based image compression methods}
Invertible Neural Networks \cite{dinh2014nice} represent a novel paradigm that ensures the reversibility of the raw data to its latent representation. It uses the change of variables formula to formulate the likelihood estimation. By using invertible modules, INNs can learn a feature representation that captures the underlying structure of the data while maintaining the ability to revert back to the original data space with no information loss.

NICE \cite{dinh2014nice} first proposed the additive coupling layer as invertible transform, which splits input dimensions into a static part and a transformed part that undergoes an additive transform. This approach provides a simple computation of the Jacobian matrix which is used in the likelihood estimation, thus enabling efficient computation and training of the model. RealNVP \cite{dinh2016density} further introduced affine coupling layers using affine function to perform the transform. Glow \cite{kingma2018glow} combined the activation normalization, invertible convolution, and an affine coupling layer together to form an invertible model capable of learning complex data distributions. INNs have been used in image generation and various image processing tasks including image super-resolution, denoising and dehazing, etc \cite{liang2021hierarchical,zhu2022high,wang2022low, liu2021invertible,li2021dehazeflow}. 

Considering INN constructs a bijection between an input image and latent feature, it is also suitable for image compression as the encoding and decoding transforms. In \cite{helminger2020lossy,wang2020modeling}, INN is explored for lossy image compression. The latent features obtained by INN are split into essential and discarded information, where the discarded information is not further encoded and replaced with noise sampling from a Gaussian distribution in the decoding process. However, the complexity of accurately modeling these features and the potential of errors in the inverse sampling process present challenges to accurate reconstruction. To address this problem, Xie et al. \cite{xie2021enhanced} proposed to eliminate the unstable sampling mechanism and use an attentive channel squeeze layer. During encoding, this layer performs an averaging operation that effectively reduces the channel dimension of the encoded representation. In the decoding, the quantized features are replicated to match the original shape. While channel squeezing reduces the channel number of features to-be-encoded, the method of simply copying features to restore the original channel number during inverse process may lead to overly smooth reconstructions, lacking the finer details and textures of the original image. Augmented Normalizing Flow (ANF)-based image compression (ANFIC) \cite{ho2021anfic} proposed to take advantage of ANF \cite{huang2020augmented} for image compression. It enhances the transform used in the coupling layer by augmenting the input space with a noise instead of splitting the input image. The existing INN-based image compression methods only take the INN as an encoder/decoder transform without investigating its effect on the image compression process. This paper is the first one that explicitly investigates the formulation of using INN for image compression, especially under the condition of quantization where the overall process is no longer invertible.

\section{Proposed Approximately Invertible Neural Network (A-INN): Formulation}

\begin{figure}[tbp]
  \centering
  \begin{subfigure}{0.4\textwidth}
    \centering
    \includegraphics[width=\textwidth]{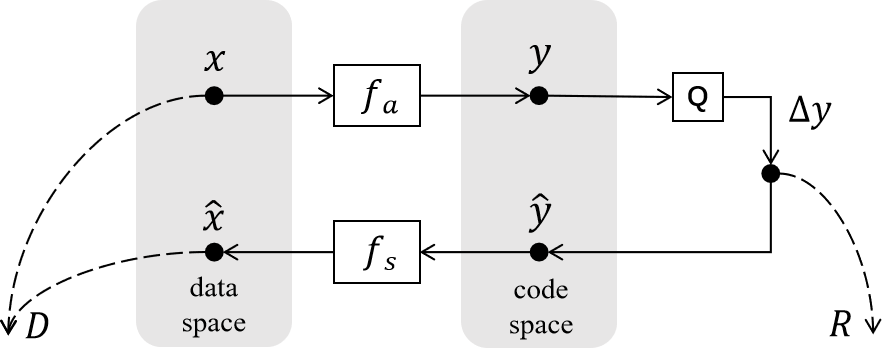}
    \abovecaptionskip=10pt 
    \belowcaptionskip=10pt 
    \caption{General VAE based learned image compression framework}
    \label{fig:E2E_IC}
  \end{subfigure}
  
  \begin{subfigure}{0.4\textwidth}
    \centering
    \includegraphics[width=\textwidth]{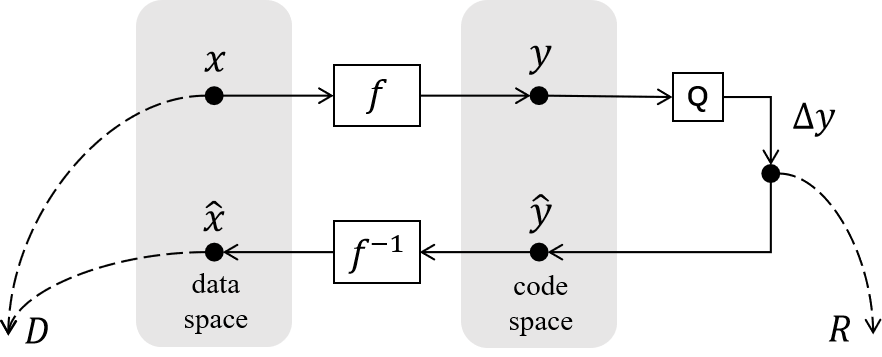}
    \abovecaptionskip=10pt 
    \belowcaptionskip=10pt 
    \caption{Invertible neural network (INN) based image compression framework}
    \label{fig:FINN_IC}
  \end{subfigure}

  \begin{subfigure}{0.4\textwidth}
    \centering
    \includegraphics[width=\textwidth]{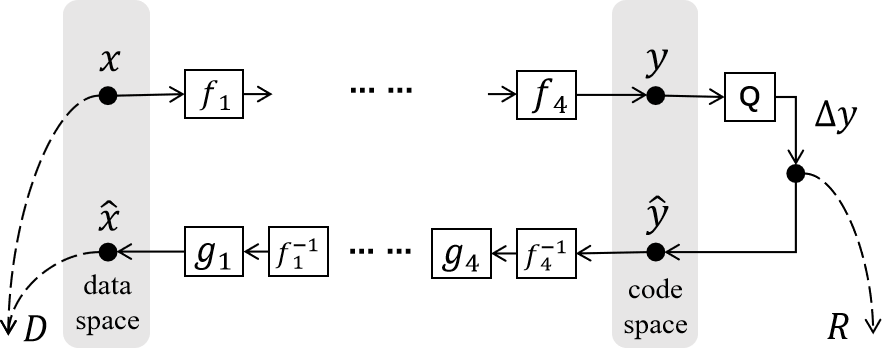}
    \abovecaptionskip=10pt 
    \belowcaptionskip=0pt 
    \caption{Proposed A-INN with progressively denoising module based image compression framework}
    \label{fig:PDM_AINN_IC}
  \end{subfigure}
  \caption{Diagram of the proposed A-INN based image compression in comparison with the existing learned image compression frameworks.}
  \label{fig:compression_comparison}
\end{figure}

The general framework of an end-to-end image compression model can be illustrated as shown in Fig. \ref{fig:E2E_IC}. It consists of an encoder (also known as the analysis transform), a decoder (also known as the synthesis transform), quantization (rounding) and entropy coding which may also include the hyperprior coding. The encoder ($f_a$) transforms the input $x$ into the hidden representation $y$, which is then rounded and entropy coded into bits, resulting in a rate $R$. The bits at the decoder are then entropy decoded and transformed by the decoder ($f_s$) into the reconstructed image $\hat{x}$, which is measured against the original input $x$, resulting in the distortion $D$ (usually in terms of MSE). The quantization is mimicked as an additive uniform noise for differentiable training, where the quantized value can be obtained as $y = y + \Delta y$.

The objective of image compression is to minimize the distortion $D$ under a rate budget $R_T$. This constrained optimization problem is usually transformed by the Lagrangian multiplier method into minimizing a weighted sum of the rate and distortion, termed as the rate-distortion (RD) cost,

  \begin{equation}
    \min J = R + \lambda D
    \label{eq:eq1}
  \end{equation}
where $\lambda$ is the Lagrangian multiplier governing the trade-off between the rate and distortion. Note that this has also been formulated as $ \min J = D + \lambda R $, which is the same with different Lagrangian multipliers.

From the variational learning perspective as in \cite{balle2018variational}, supposing the input follows the distribution $x \sim p(x)$, the above rate-distortion optimization problem over the whole input data can be formulated by

  \begin{equation}
    \min \mathbb{E}_{x \sim p(x)} \mathbb{E}_{\hat{y} \sim q(\hat{y}|x)}(R + \lambda D)
    \label{eq:eq2}
  \end{equation}
where $\hat{y} \sim q(\hat{y}|x)$ is the uniform distribution of the noise added in the quantization process.

With a flow-based invertible neural network as the backbone of the image compression network as shown in Fig. \ref{fig:FINN_IC}, the encoder and decoder are coupled as invertible transforms ($f$ and $f^{-1}$). In such a case, the reconstruction error can be obtained as
  \begin{equation}
    \begin{aligned}
        D &= \| x - \hat{x} \| \\
          &= \| f^{-1}(\hat{y}) - f^{-1}(y) \| \\
          &= \| f^{-1}(y + \Delta y) - f^{-1}(y) \|
    \end{aligned}
    \label{eq:eq3}    
  \end{equation}
where $x = f^{-1}(y)$ is the inverse of the encoder transform. $\|\cdot\|$ represents the norm used to calculate the distortion, which is usually the $L_2$ norm corresponding to the MSE loss.
  
Using Taylor's theorem, the first term $f^{-1}(y + \Delta y)$ can be further expanded as
  \begin{equation}
    f^{-1}(y + \Delta y) = f^{-1}(y) + \frac{\partial f^{-1}(y)}{\partial y} \Delta y + o(\Delta y)
  \label{eq:eq4}
  \end{equation}
where $\frac{\partial f^{-1}(y)}{\partial y}$ is the Jacobian matrix of $f^{-1}(y)$. $o(\Delta y)$ are the higher-order terms which can be omitted. Thus, Eq. \ref{eq:eq3} can be transformed into
\begin{small}
{
  \begin{equation}
    D \approx \left\| f^{-1}(y) + \frac{\partial f^{-1}(y)}{\partial y} \Delta y - f^{-1}(y) \right\| = \left\| \frac{\partial f^{-1}(y)}{\partial y} \Delta y \right\|
  \label{eq:eq5}
  \end{equation}
}
\end{small}

According to the compatibility theorem of matrix norm and vector norm, it can be further transformed to:
  \begin{equation}
    D \leq \left\| \frac{\partial f^{-1}(y)}{\partial y} \right\| \left\| \Delta y \right\|
  \label{eq:eq6}
  \end{equation} 
which is the upper bound of the distortion, and minimizing this bound results in minimizing the distortion.

On the other hand, the rate is formulated as the entropy which is $-\log(p(\hat{y}))$. So by combining the rate term and the distortion term in Eq. \ref{eq:eq6}, the optimization objective in Eq. \ref{eq:eq2} can be expressed as:
\begin{small}
{  
  \begin{equation}
    \min \mathbb{E}_{x \sim p(x)} \mathbb{E}_{\hat{y} \sim q(\hat{y}|x)} \left(-\log(p(\hat{y})) + \lambda \left\| \frac{\partial f^{-1}(y)}{\partial y} \right\| \left\| \Delta y \right\| \right)
  \label{eq:eq7}
  \end{equation}
}  
\end{small}

With $q(\hat{y}|x)$ following a uniform distribution, Eq. \ref{eq:eq7} can be further simplified into:
  \begin{equation}
    \min \mathbb{E}_{x \sim p(x), \Delta} \left( -\log(p(\hat{y})) + \lambda \cdot n_q \left\| \frac{\partial f^{-1}(y)}{\partial y} \right\| \right)
  \label{eq:eq8}
  \end{equation}
where $\mathbb{E}_{x \sim p(x), \Delta}$ represents the input $x$ being extensively sampled to generate uniformly distributed $\hat{y}$ in the learning process. $n_q$ represents the expectation over the norm of the uniformly distributed noise $U(-0.5, 0.5)$, which equals 0.25.

In contrast, the formulation of the invertible neural network for generative modelling as provided in \cite{dinh2014nice} can be expressed over the input $x \sim p(x)$ as

  \begin{equation}
    \max \mathbb{E}_{x \sim p(x)} [\log(p(y)) + \log(|\det(\frac{\partial f(x)}{\partial x})|)]
    \label{eq:eq9}
  \end{equation}  
which can be turned into:
  \begin{equation}
      \begin{aligned}
          \min \mathbb{E}_{x \sim p(x)} [-\log(p(y)) - \log(|\det(\frac{\partial f(x)}{\partial x})|)] \\
          = \mathbb{E}_{x \sim p(x)} [-(\log(p(y)) + \log(|\det(\frac{\partial f^{-1}(y)}{\partial y})|))]
      \end{aligned}
  \label{eq:eq10}
  \end{equation}
where the determinant of the forward transform $\det(\frac{\partial f(x)}{\partial x})$ equals the inverse of the determinant of the invertible transform $1 / {\det(\frac{\partial f^{-1}(y)}{\partial y})}$.

\begin{figure*}[tbp]
	\centering
	\includegraphics[width=1.0\hsize]{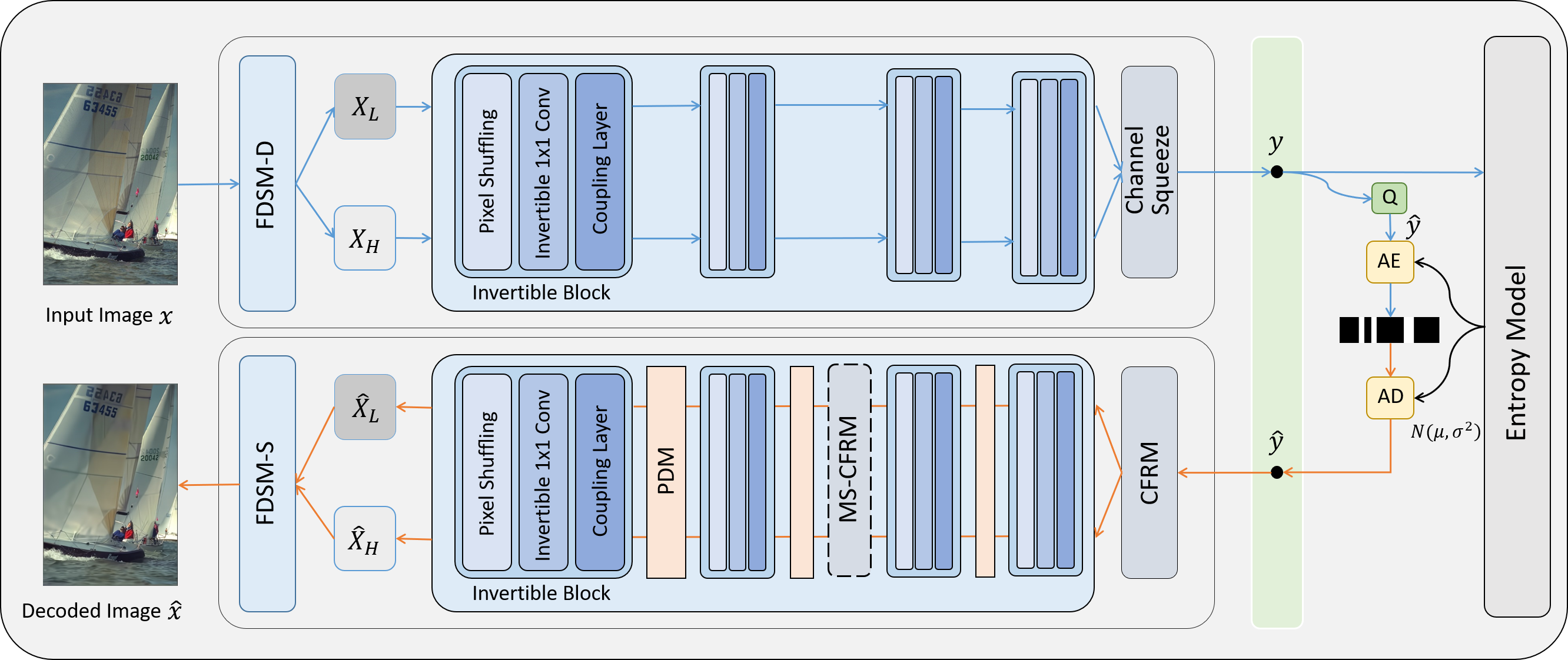}
	\caption{Overview of the proposed Approximately Invertible Neural Network (A-INN) based image compression framework. It includes the invertible block as the main transform, the proposed FDSM containing the frequency enchanced decomposition (FDSM-D) and synthesis (FDSM-S), PDM for denoising due to quantization, CFRM and the optional MS-CFRM for denoising due to channel squeeze. Q represents quantization, AE and AD represent the arithmetic encoder and arithmetic decoder, respectively, and Entropy Model represents the joint hyperprior and context based model. }
  \label{fig:framework}
\end{figure*}

By comparing the invertible formulation in Eq. \ref{eq:eq10} and our formulation of the invertible neural network with quantization error in Eq. \ref{eq:eq8}, it can be seen that while they share a similar form with the entropy (rate) and manipulation of the Jacobian matrix, they differ in the specific processing of the Jacobian matrix, namely the determinant and norm, respectively. We term our formulation of the invertible neural network with quantization error as Approximately Invertible Neural Network (A-INN).

A-INN is different from the INN from two perspectives. Firstly, the INN takes the determinant of the Jacobian matrix in the optimization aiming to represent the data distribution by penalizing contraction and encouraging expansion in regions of high density of the data. By contrast, A-INN uses the norm of the Jacobian matrix in the optimization aiming to transform the data into an error-resilient feature space in order to reduce the effect of the quantization noise. In a word, the learning objective of the transform network in INN and A-INN is different. Secondly, A-INN is controlled by the Lagrangian multiplier $\lambda$ to balance the rate and the reconstruction quality. On the contrary, INN is to learn the projection between a prior distribution and the data distribution where the entropy (rate) term and the determinant of the Jacobian matrix term represent the balance between contracting the data to features following the prior distribution and expanding the features to represent meaningful structures in the data.


In addition to illustrating the difference between the INN and A-INN, the above formulation also provides a new avenue to the optimization of A-INN. According to Eqs. \ref{eq:eq6} and \ref{eq:eq7}, it can be seen that minimizing the distortion can be achieved by minimizing both $\left\| \frac{\partial f^{-1}(y)}{\partial y} \right\|$ and $\left\| \Delta y \right\|$, where the $\left\| \frac{\partial f^{-1}(y)}{\partial y} \right\|$ is the norm of the Jacobian matrix and $\left\| \Delta y \right\|$ represents the norm of the uniform noise applied during training. Since $y$ and $\Delta y$ are independent from each other, the two terms can also be minimized individually. Specifically, minimizing $\left\| \frac{\partial f^{-1}(y)}{\partial y} \right\|$ makes the network learn an error-resilient transform, which can be done by designing better invertible architectures. On the other hand, minimizing $\left\| \Delta y \right\|$ can also lead to the minimization of the distortion, which has been ignored in previous works. In this paper, we propose a progressively denoising module in the approximately invertible neural network (A-INN) by reducing the added noise in the flow-based invertible model. Specifically, minimizing $\left\| \Delta y \right\| = \left\| \hat{y} - y \right\|$ can be performed as a denoising process to reduce the noise from the feature. Based on this, a denoising function $g$ is designed and inserted into the invertible model to reduce the noise in the feature in addition to the coupled invertible transform $f$ and $f^{-1}$. The invertible network is first decomposed into progressive invertible modules (taking four sub-invertible modules for example $f = f_1 \cdot f_2 \cdot f_3 \cdot f_4$) and denoising functions ($g_1, g_2, g_3, g_4$) can be inserted in-between two sub-invertible modules as shown in Fig. \ref{fig:PDM_AINN_IC}. 

The above A-INN formulation not only can be applied to lossy image compression, but also can be extended to any INN based lossy compression method as theoretical foundation. It also distinguishes INN based lossy compression from the original INN based image generative modelling, and paves a new avenue to the optimization of INN.

\section{Proposed Approximately Invertible Neural Network (A-INN): Architecture}
\subsection{A-INN with Progressive Denoising Modules}
While this paper focuses on the formulation of A-INN for lossy image compression, a detailed A-INN architecture has also been developed for learned image compression. The framework of the proposed A-INN architecture is shown in Fig. \ref{fig:framework}. It is based on a variational autoencoder architecture using invertible modules as the main component for encoder and decoder as in \cite{xie2021enhanced}. Three modules are proposed on top of the backbone, including the Progressive Denoising Modules (PDM), the Cascaded Feature Recovery Module (CFRM), and the Frequency-enhanced Decomposition and Synthesis Module (FDSM).

In the encoding stage, the frequency-enhanced decomposition module is used first, which separates the input image \( x \) into high-frequency enhanced component \( x_H \) and low-frequency enhanced component \( x_L \) components. These components are combined and then processed by an invertible transformation block to enable interaction between high and low frequencies. Subsequently, an attentive channel squeeze layer is applied to reduce the channel dimensions of the features. The features are then quantized. Additive uniform noise is used to emulate the quantization effect during training, while quantization is directly used during testing. Regarding the entropy model, to maintain consistency and facilitate performance comparison with the baseline model \cite{xie2021enhanced}, the joint hyperprior and context based single Gaussian entropy model \cite{minnen2018joint} is used to formulate the distribution of the latent feature and arithmetic coding is then used for encoding.

In the decoding phase, the entropy-decoded features are first processed by the CFRM to restore their original channel dimensions. These features are then passed through the inverse of the invertible transformation block. Between these inverse transformation blocks, Progressive Denoising Modules (PDM) are integrated to reduce the quantization noise. Finally, a frequency-enhanced synthesis module reconstructs the image, resulting in the output \( \hat{x} \).

The invertible transformation block used in the encoder and decoder typically consists of three main components: pixel shuffling, invertible \( 1 \times 1 \) convolution and coupling layer. First, a pixel shuffling layer is employed to halve the spatial resolution of the input tensor, while concurrently amplifying the channel dimension by a factor of four. Then, an invertible \( 1 \times 1 \) convolution provides an invertible and learnable processing of the channel dimensions, which enhances the interaction of features. Finally, coupling layer typically splits the input features into two parts along the channel axis and apply affine transformations to couple them, enabling the model to learn more efficient feature representations. It is worth noting that the used coupling layer diverges from the conventional one. With the proposed FDSM module (with details in subsection \ref{subsec:fdsm}), an image is split into high-frequency and low-frequency components. The coupling layer leverages the high- and low-frequency enhanced components \( (x_H, x_L) \) for affine coupling and interaction, with the objective of integrating high- and low-frequency information. The coupling layer used in the proposed method can be represented as follows:
\begin{align}
  y_L &= x_L \odot \exp{\left(\sigma\left(f_{\theta1}\left(x_H\right)\right)\right)} + g_{\phi1}\left(x_H\right)\\
  y_H &= x_H \odot \exp{\left(\sigma\left(f_{\theta2}\left(y_L\right)\right)\right)} + g_{\phi2}\left(y_L\right)
\end{align}

\begin{figure}[tbp]
	\centering
	\includegraphics[width=0.4\textwidth]{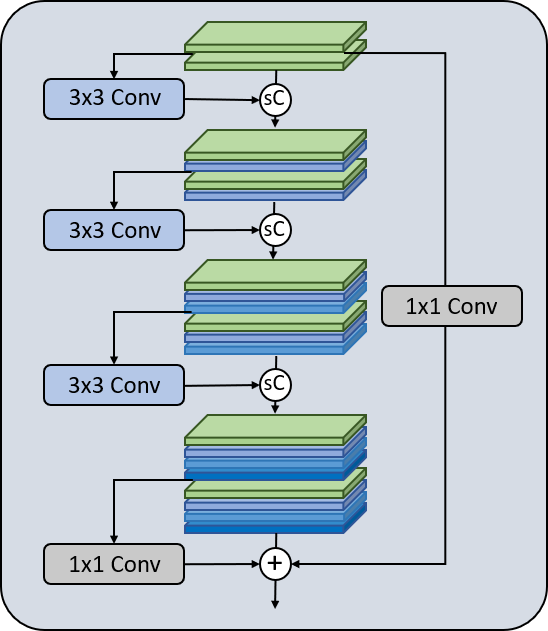}
	\caption{Proposed Cascaded Feature Recovery Module (CFRM). ReLU is used after the $3\times 3$ convolution while $1\times 1$ convolution is only used for channel processing and  no activation is used. $sC$ represents the split concatenation to concatenate the feature channels by group corresponding to the channel squeeze at the encoder side. }
  \label{fig:CFRM}
\end{figure}

The reverse process of the affine coupling layers is as follows:
\begin{align}
  x_H &= \left(y_H - g_{\phi2}\left(y_L\right)\right) / \exp{\left(\sigma\left(f_{\theta2}\left(y_L\right)\right)\right)}\\
  x_L &= \left(y_L - g_{\phi1}\left(x_H\right)\right) / \exp{\left(\sigma\left(f_{\theta1}\left(x_H\right)\right)\right)}
\end{align}
where \( \odot \) denotes the Hadamard product, \( \exp{\left(\cdot\right)} \) denotes the exponential function, and \( \sigma \) represents the Sigmoid activation function. The mapping functions \( f_{\theta1}, g_{\phi1}, f_{\theta2}, \) and \( g_{\phi2} \) can be implemented by any neural network. Here, the same network implementation as in \cite{xie2021enhanced} is used. \( x_H \) and \( x_L \) represent the input high- and low-frequency enhanced components, respectively, while \( y_H \) and \( y_L \) denote the high and low-frequency enhanced components output by the affine coupling layer.

As analyzed in the above Formulation subsection, the invertible decoder can be designed as separate invertible functions inserted with denoising functions (\( g_1, g_2, g_3, g_4 \)) to reduce the noise introduced in the quantization, to progressively achieve approximately invertible neural network (A-INN). It is crucial to select an appropriate denoising network architecture to collaborate with the invertible function in order to better reduce noise. Three architectures have been used for denoising and tested in the experiments, including the residual block-based denoising, the dense block-based denoising, and the structure-based denoising \cite{he2016deep,huang2017densely,ronneberger2015u}. For the residual block-based denoising, two residual blocks are used, where each layer includes a shortcut connection and two layers of \( 3 \times 3 \) convolution with ReLU activation. The dense block-based denoising network contains a dense layer, three convolutional layers with kernel size \( 3 \times 3 \), \( 1 \times 1 \), and \( 3 \times 3 \) to reduce the channels, and ending with another dense layer. As described in the above Formulation subsection, the noise can be removed by formulating the original feature. Therefore, the structure of the feature is essential for denoising the invertible features, and thus a structure-based denoising module is developed with down-sampling to extract global structure and guide the denoising in a similar way as the U-Net architecture. It starts with a \( 3 \times 3 \) convolutional layer, followed by a symmetrically arranged downsampling and upsampling path interconnected by skip connections, and concludes with a \( 1 \times 1 \) convolutional layer. Ablation study shown in the Experiments section verifies that all progressive denoising modules improve the performance while the structure-based denoising achieves the best, agreeing with the above analysis.

\subsection{Cascaded Feature Recovery Module}

In the encoder of the invertible transform, the spatial resolution of the image/feature is gradually reduced while the number of channels is correspondingly increased due to the pixel shuffling in the invertible structure. For example, a \(2 \times\) spatial down-shuffling results in a \(4 \times\) channel increasing, and the overall invertible encoder brings a significant increase in feature channels. The existing methods mostly take the invertible encoder as a whole and only reduce the channel dimension at the end of the invertible encoder by averaging over a certain number of channels (\(\alpha = 6\) for example). At the decoder side, the features need to be restored to the dimension before channel reduction, which is done by duplicating \(\alpha\) times. However, this replication may result in excessively smooth reconstructions, lacking the finer details and textures. Moreover, this process can also be regarded as a noise injection process to the encoder features, and thus needs to be carefully addressed in the decoder.

To overcome this problem, a Cascaded Feature Recovery Module (CFRM) as shown in Fig. \ref{fig:CFRM} is developed, which plays a pivotal role in not only restoring the dimensionality of the decoded features but also in improving the reconstructed features. The CFRM weaves together a series of carefully crafted convolutional layers, feature concatenation operations and different types of connections including residual and dense connections. It expands the dimensionality of the decoded features from \(\left(\frac{H}{s} \times \frac{W}{s} \times N\right)\) to \(\left(\frac{H}{s} \times \frac{W}{s} \times \alpha N\right)\), where \(N\) and \(\alpha\) represents the number of channels in the latent features and the increase times of the channels, respectively, and \(s\) represents the spatial down-sampling. This enriches the feature information and provides a more detailed representation of the input.

Specifically, a \(3 \times 3\) convolutional layer with ReLU activation function is used to process the input features and is concatenated with the input features as a form of dense connection and also increase the number of channels. This dense processing is repeated for another three times to obtain the desired increase of channels in the features. This cascading process can be summarized by the following formula:
{
  \begin{equation}
      {\hat{y}}_l = s\mathbb{C}\left({\hat{y}}_{l-1}, \ \sigma \left({\mathrm{Conv}}_{3\times3}\left({\hat{y}}_{l-1}\right)\right)\right),\quad l=1,2,3
  \end{equation}
}
where \(s\mathbb{C}\) denotes the feature split concatenation operation to concatenate the feature channels by group corresponding to the channel squeeze at the encoder side. \(\text{Conv}_{3 \times 3}\) represents the \(3 \times 3\) convolutional layer with $\sigma$ representing the ReLU activation function, \(\hat{y_0}\) and \(\hat{y_l}\) represents the input feature to the CFRM module and the intermediate features, respectively.

To further enhance the quality of the decoded features, a \(1 \times 1\) convolutional layer is applied to both the original input features and the output feature \(y_3\), and added together to obtain the final feature. This process can be represented as:
  \begin{equation}
      {\hat{y}}_{out} = {\mathrm{Conv}}_{1\times1}\left({\hat{y}}_0\right) + {\mathrm{Conv}}_{1\times1}\left({\hat{y}}_3\right)
  \end{equation}
where \(\hat{y}_{\text{out}}\) represents the output feature. The designed CFRM not only increases the number of features but also learns to enhance the quality of the output feature. The cascading nature of the CFRM facilitates a smooth transition of information across channels and also provides a richer feature representation than the duplicated representation, which is useful to maintain coherence and continuity in the reconstructed image.

Moreover, a Multi-Stage Cascaded Feature Recovery Module (MS-CFRM) is further developed that incrementally restores half of the feature channels in each stage. Specifically, the first stage of recovery is implemented subsequent to the completion of the second invertible transformation block, while the second stage of recovery is conducted after the fourth invertible transformation block. In this way, the complexity of the feature processing at the later invertible blocks can be reduced since the number of feature channels is reduced. Also this can further explore the invertible block to be gradually more noise-resilient to deal with the channel reduction.

\subsection{Frequency-enhanced Decomposition and Synthesis Module}
\label{subsec:fdsm}

It is known that high-frequency information is relatively difficult to be restored through a neural network. To better preserve the high-frequency information, a Frequency-enhanced Decomposition and Synthesis Module (FDSM) is developed to explicitly enhance the high-frequency information contained in the original image. It consists of a frequency-enhanced decomposition module as shown in Fig. \ref{fig:FEDM}, and a corresponding dual-frequency attention based synthesis module as shown in Fig. \ref{fig:DFASM}. This decomposition and synthesis strategy allows us to independently operate on the high- and low-frequency enhanced components, and then recombine these processed components to generate the final image.

\subsubsection{Frequency-enhanced Decomposition Module}

\begin{figure}[tbp]
  \centering
  \begin{subfigure}{0.495\textwidth}
    \centering
    \includegraphics[width=\textwidth]{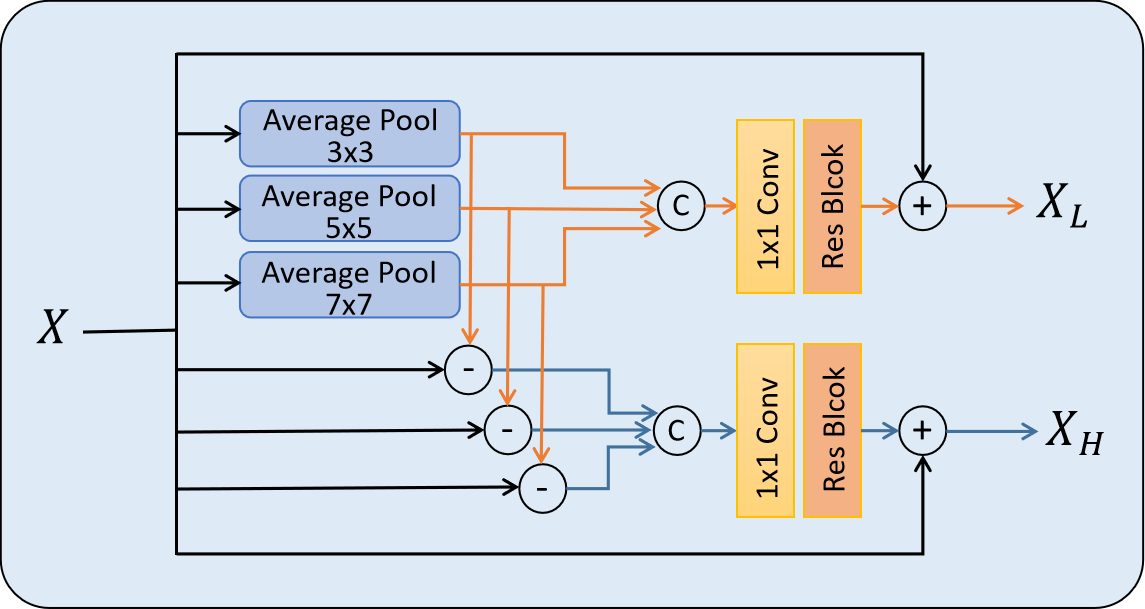}
    \abovecaptionskip=10pt 
    \belowcaptionskip=10pt 
    \caption{Frequency-enhanced decomposition (FDSM-D).}
    \label{fig:FEDM}
  \end{subfigure}
  
  \begin{subfigure}{0.495\textwidth}
    \centering
    \includegraphics[width=\textwidth]{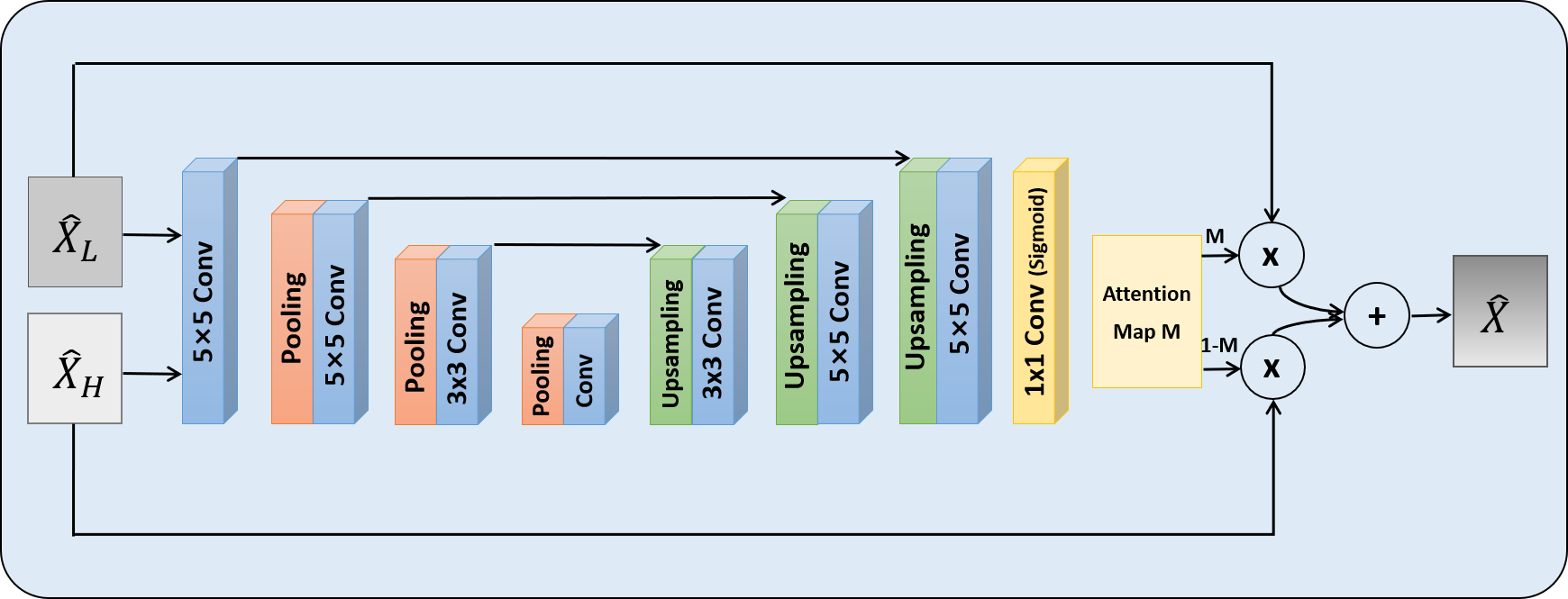}
    \abovecaptionskip=10pt 
    \belowcaptionskip=0pt 
    \caption{Dual-frequency attention based synthesis (FDSM-S).}
    \label{fig:DFASM}
  \end{subfigure}
  \caption{Frequency-enhanced Decomposition and Synthesis Module (FDSM). ReLU is used after the convolution block except the $1\times 1$ convolution used in the end of synthesis which uses Sigmoid as activation to produce the attention map.}
  \label{fig:FEDSM}
\end{figure}

In the decomposition, the high-frequency information is extracted by subtracting the low-frequency information from the original image as in Fig. \ref{fig:FEDM}. As shown in Fig. \ref{fig:FEDM}, three average pooling operations of different sizes (\( 3 \times 3 \), \( 5 \times 5 \), \( 7 \times 7 \)) are used to perform local averaging on the image, thereby extracting low-frequency information (\(X_{L_i}, i=1,2,3\)) at different scales. Subsequently, by subtracting the low-frequency information \(X_{L_i}\) from the original image \(X\), the corresponding high-frequency information can be obtained (\(X_{H_i}=X-X_{L_i}, \quad i=1,2,3\)). Then the different scales of low-frequency information and high-frequency information are fused together, respectively, by first applying a \(1\times1\) convolutional layer to reduce the number of channels and a residual block to process the features. Finally, the low- and high-frequency features are added to the original image to obtain the final low- and high-frequency enhanced inputs, respectively.

\subsubsection{Dual-Frequency Attention based Synthesis Module}
The low- and high-frequency enhanced inputs are then encoded and decoded with the proposed A-INN, where the outputs are the reconstructed low- and high-frequency enhanced information due to the invertible nature of INN. A Dual-Frequency Attention based Synthesis Module is developed to synthesize the final output image by fusing the reconstructed low- and high-frequency information. It is designed based on an attention mechanism where the attention is applied on the dual frequency information, as shown in Fig. \ref{fig:DFASM}. It learns a pixel-level attention weight to achieve precise fusion of different frequency information. The attention weight is obtained using a UNet to effectively combine the dual frequency information at multiple scales. A \(1\times1\) convolutional layer with Sigmoid activation function is used to process the output of UNet to compute attention weights \(M\) for each pixel. The overall attention estimation process can be formulated as:

  \begin{equation}
      M = \sigma_s\left(\mathrm{Conv}_{1\times1}\left(f_{att}\left(\mathbb{C}\left[{\hat{X}}_L,{\hat{X}}_H\right]\right)\right)\right)
  \end{equation}
where \( f_{att} \) represents the UNet function and \( \sigma_s \) represents the Sigmoid activation function.

Subsequently, the weights \( M \) and \( 1-M \) are applied to the high-frequency and low-frequency components, respectively, and then added together to generate the final output image, corresponding to the frequency-enhanced decomposition. The reconstructed image \( \hat{X} \) can be represented by:

  \begin{equation}
      \hat{X} = \hat{X}_L \cdot M + \hat{X}_H \cdot (1-M)     
  \end{equation}
where \( \hat{X}_L \) and \( \hat{X}_H \) represent the reconstructed low- and high-frequency enhanced component, respectively. The Dual-Frequency Attention based Synthesis Module effectively integrates features of different frequencies, producing a reconstructed image rich in detail and of high-frequency information.

\begin{figure*}[tbp]
  \centering
  \begin{subfigure}{0.245\textwidth} 
    \centering
    \adjustbox{valign=m}{\includegraphics[width=\linewidth]{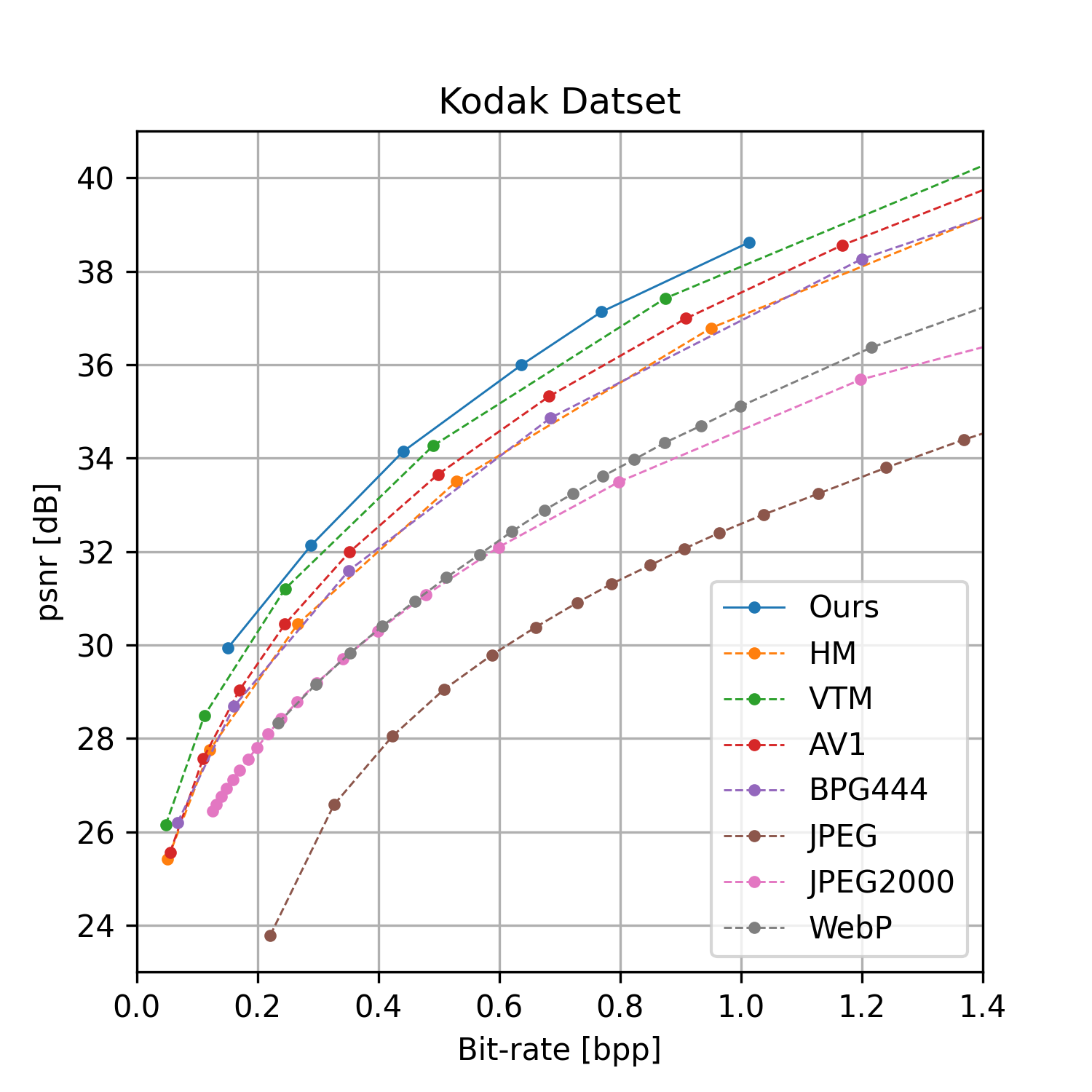}}
    \caption*{(a)} 
  \end{subfigure}
  \hfill
  \begin{subfigure}{0.245\textwidth} 
    \centering
    \adjustbox{valign=m}{\includegraphics[width=\linewidth]{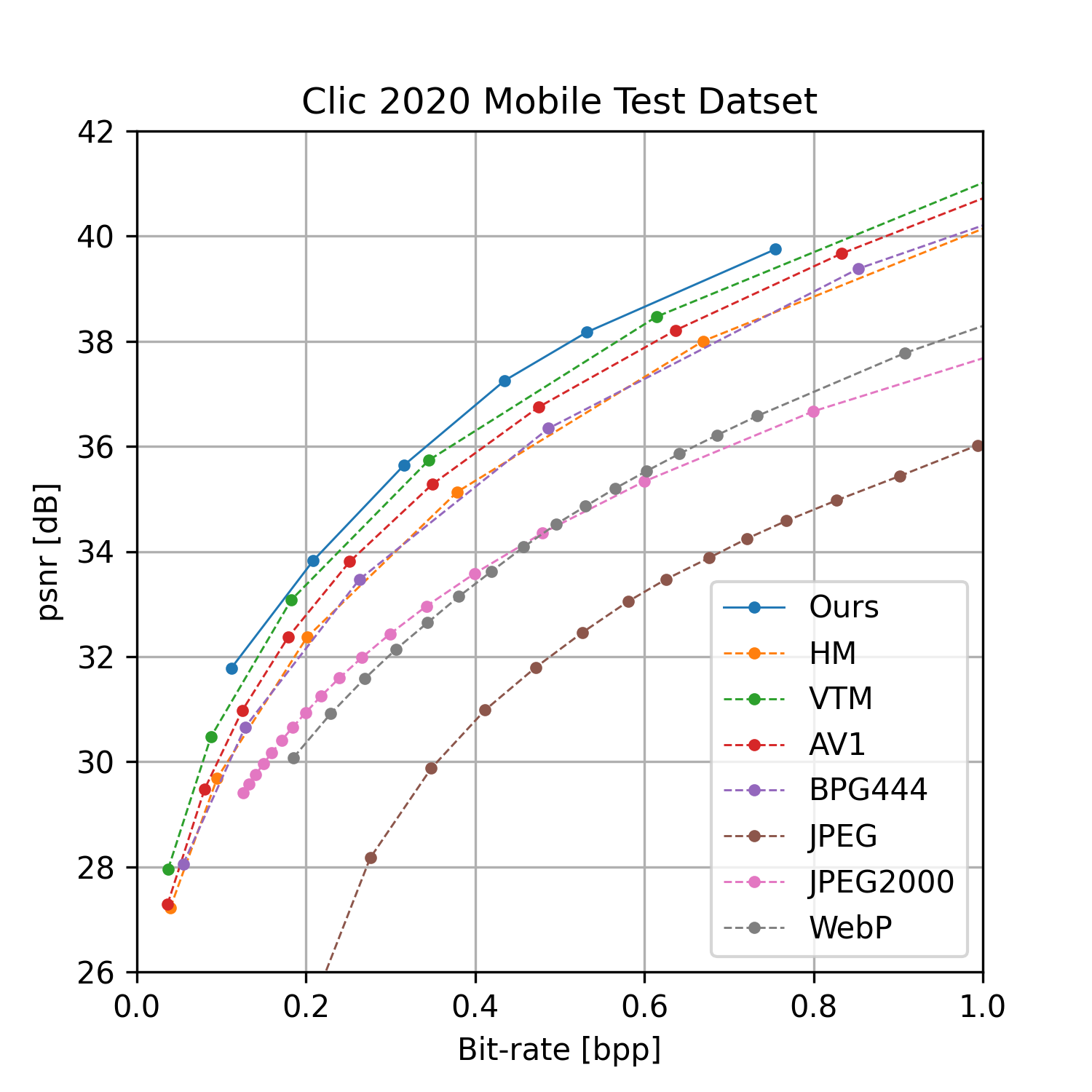}}
    \caption*{(b)} 
  \end{subfigure}
  \hfill
  \begin{subfigure}{0.245\textwidth} 
    \centering
    \adjustbox{valign=m}{\includegraphics[width=\linewidth]{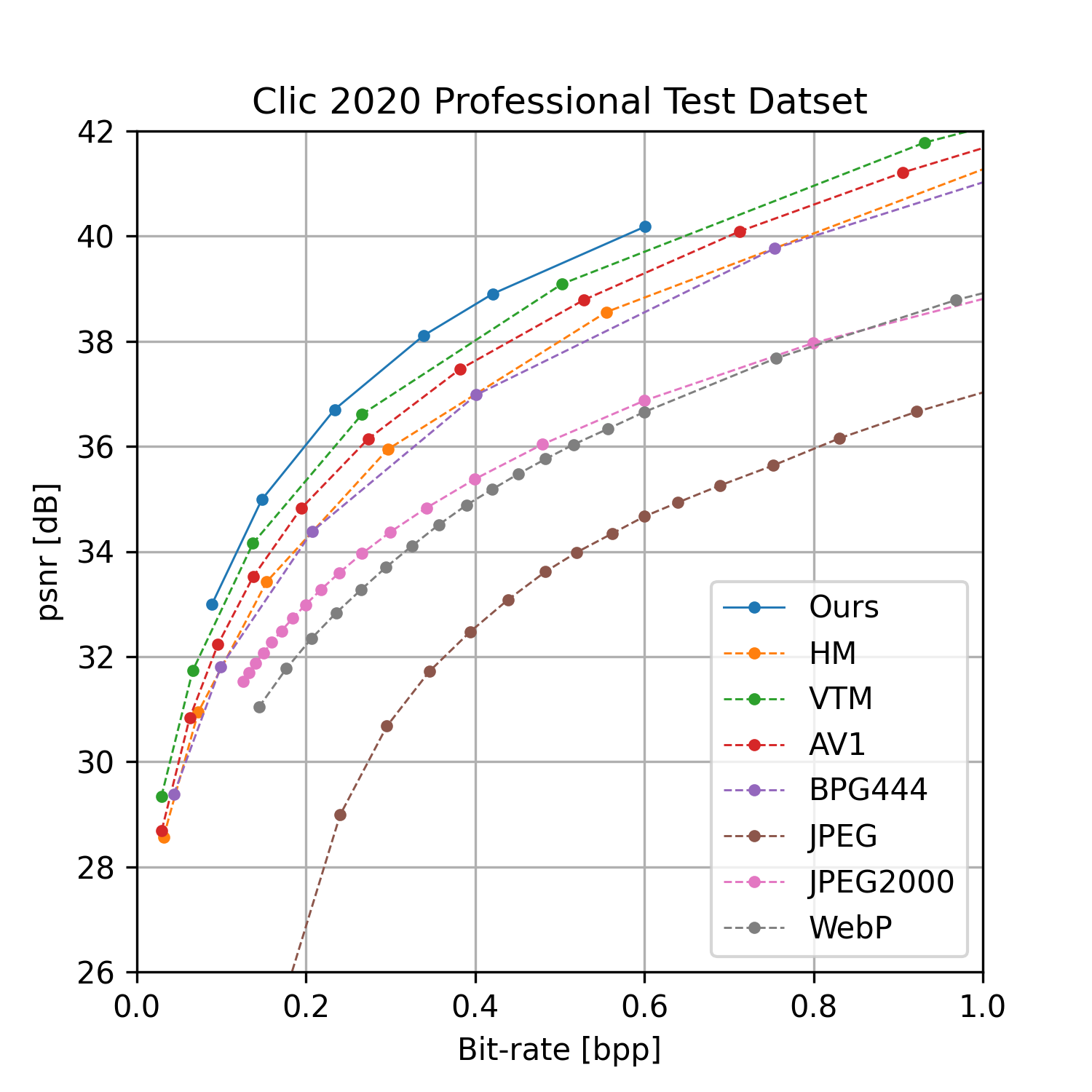}}
    \caption*{(c)} 
  \end{subfigure}
  \hfill
  \begin{subfigure}{0.245\textwidth} 
    \centering
    \adjustbox{valign=m}{\includegraphics[width=\linewidth]{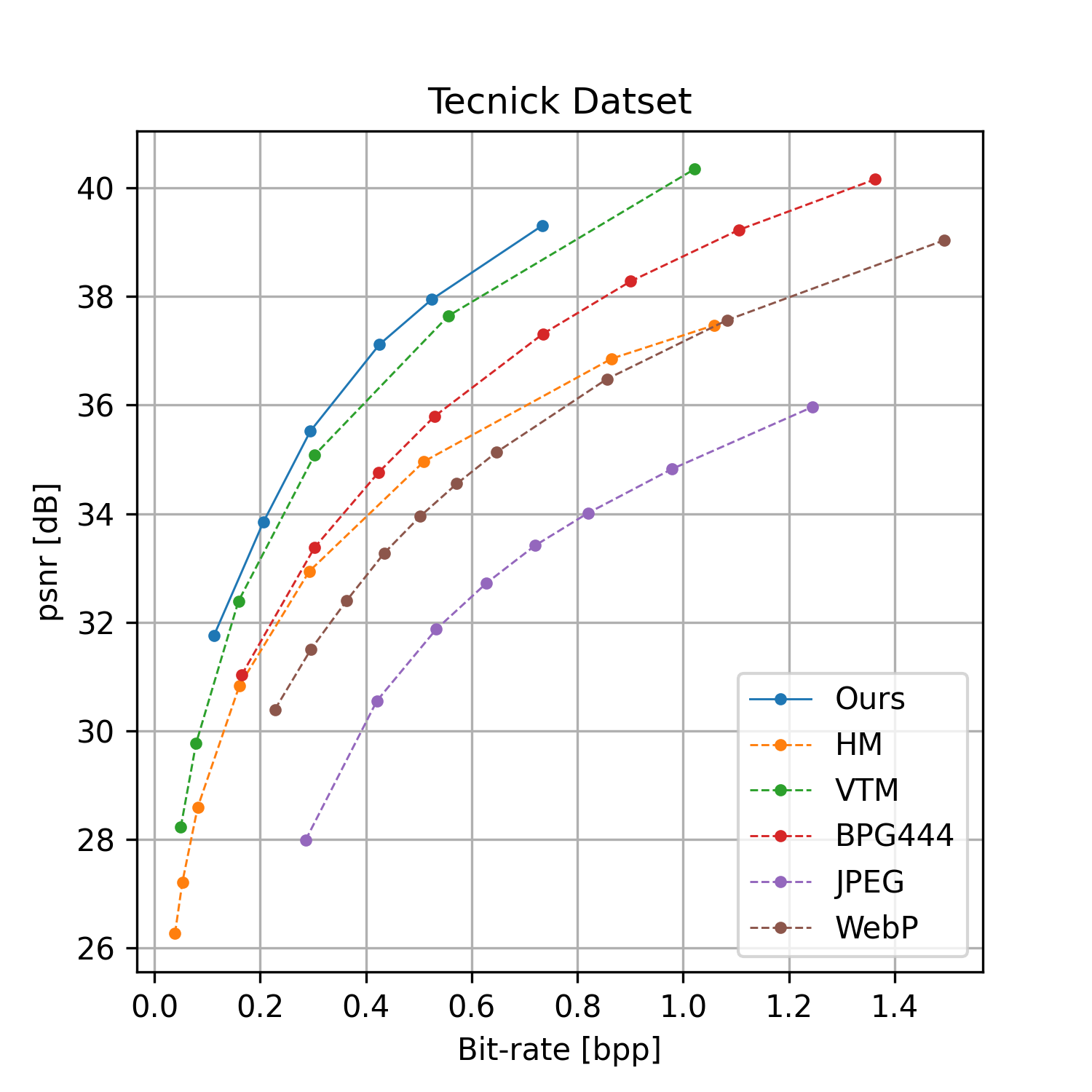}}
    \caption*{(d)} 
  \end{subfigure}

  \begin{subfigure}{0.245\textwidth} 
    \centering
    \adjustbox{valign=m}{\includegraphics[width=\linewidth]{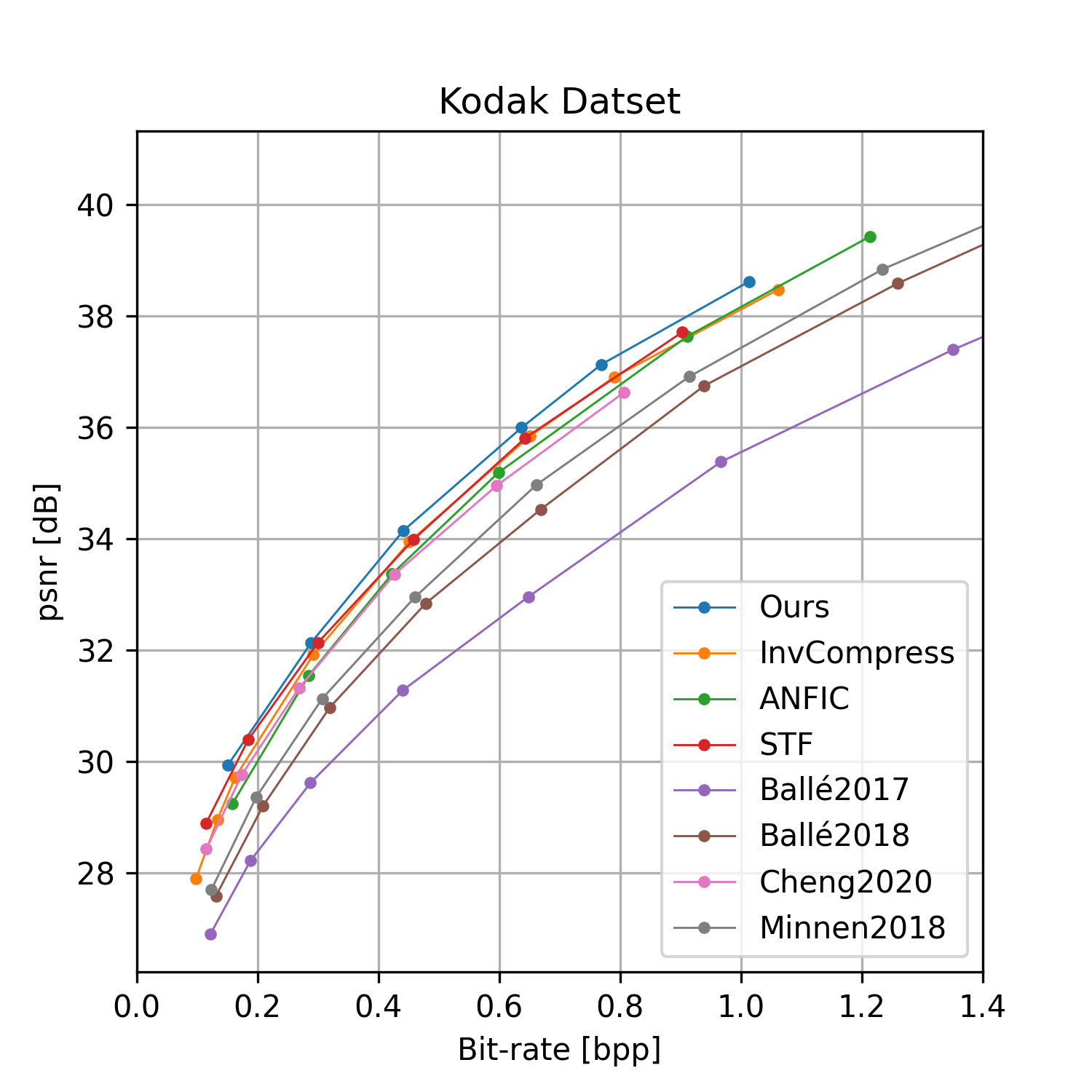}}
    \caption*{(e)} 
  \end{subfigure}
  \hfill
  \begin{subfigure}{0.245\textwidth} 
    \centering
    \adjustbox{valign=m}{\includegraphics[width=\linewidth]{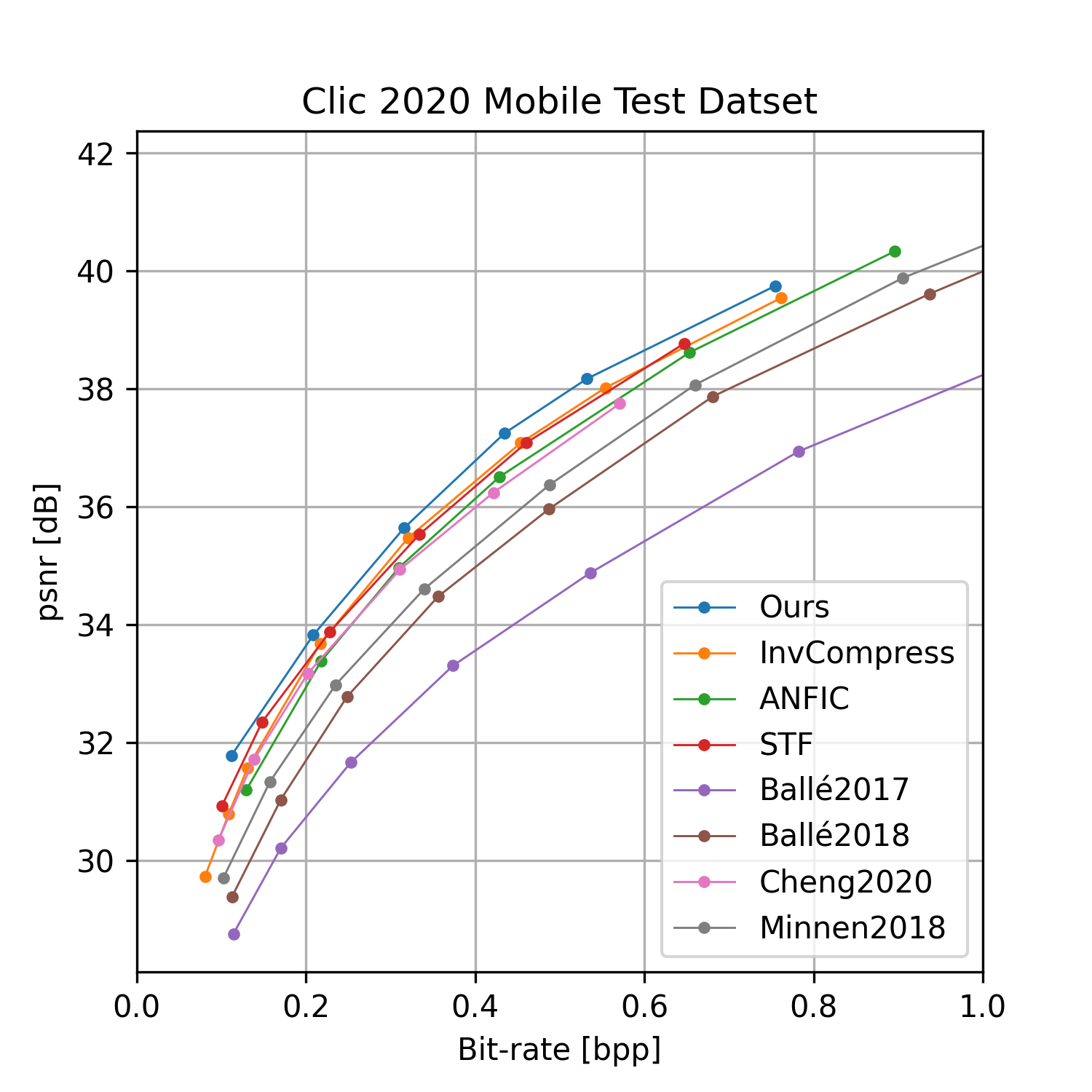}}
    \caption*{(f)} 
  \end{subfigure}
  \hfill
  \begin{subfigure}{0.245\textwidth} 
    \centering
    \adjustbox{valign=m}{\includegraphics[width=\linewidth]{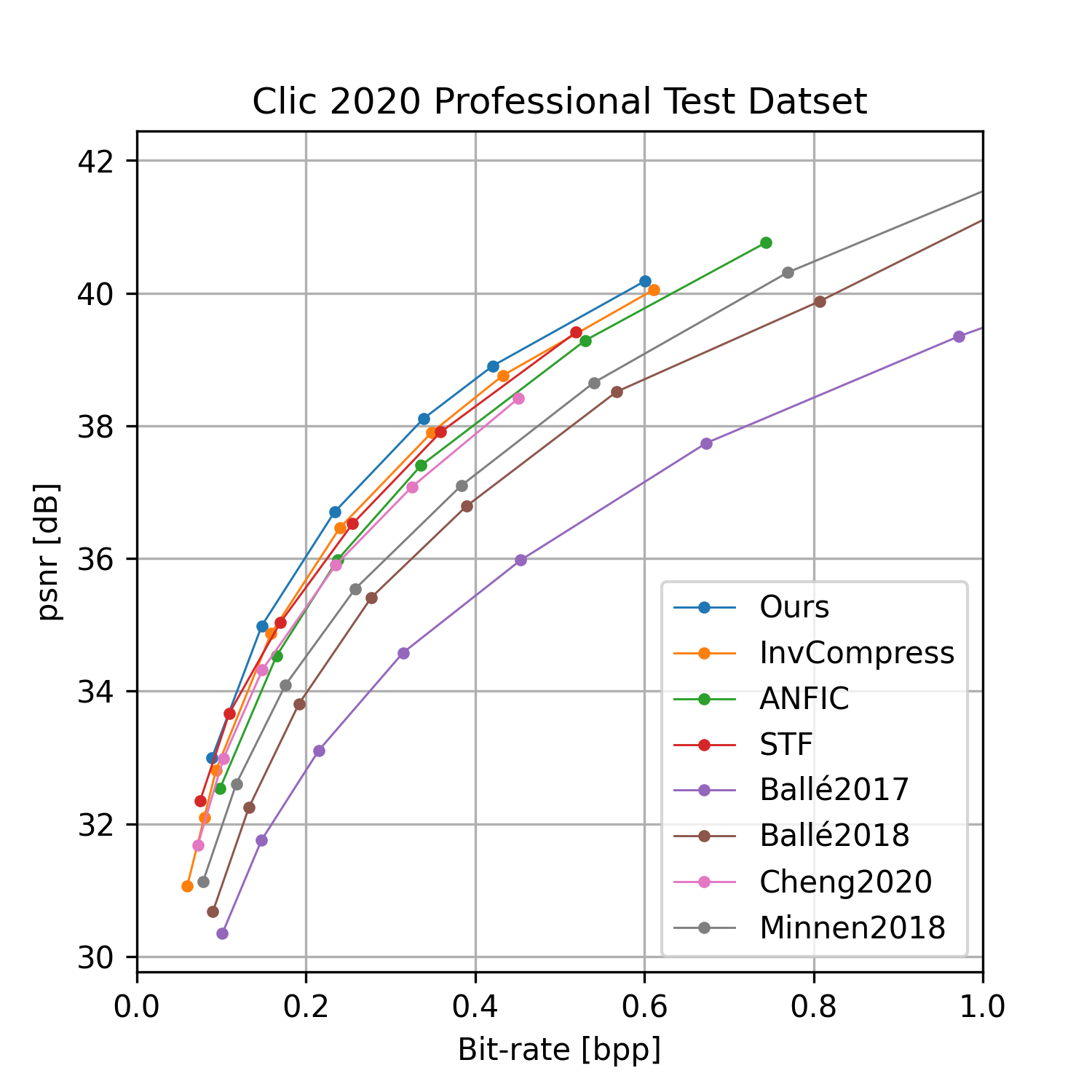}}
    \caption*{(g)} 
  \end{subfigure}
  \hfill
  \begin{subfigure}{0.245\textwidth} 
    \centering
    \adjustbox{valign=m}{\includegraphics[width=\linewidth]{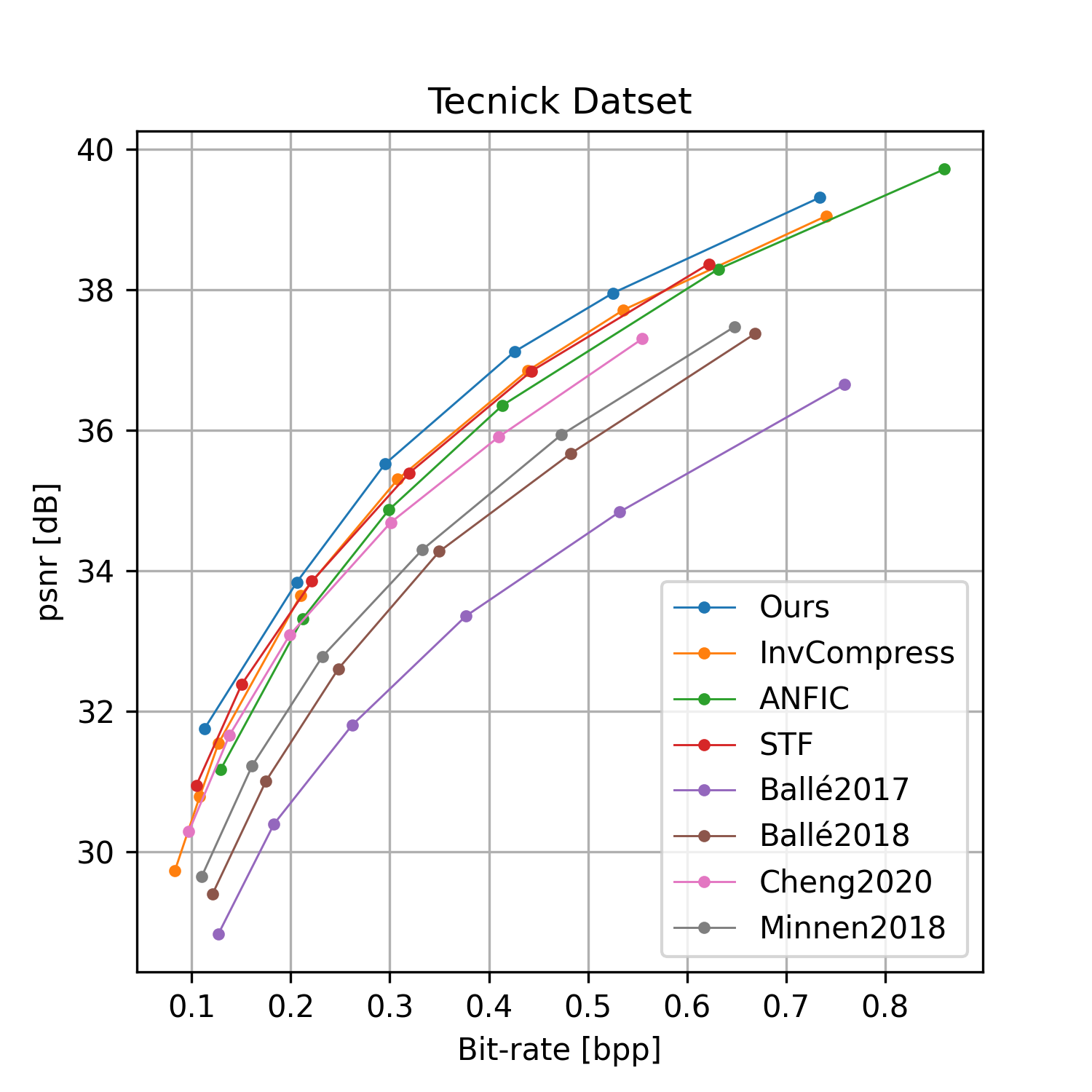}}
    \caption*{(h)} 
  \end{subfigure}
  \caption{RD curve comparison of the proposed A-INN against the existing methods, including the traditional methods (first row (a)-(d)) and the learned image compression methods (second row (e)-(h)), on Kodak, CLIC Mobile, CLIC Professional and Tecnick dataset.} 
  \label{fig:rd-curv}
\end{figure*}

\vspace{0.2cm}
\section{Experiments}
\subsection{Experimental setup}
\textbf{Training details:} The CompressAI PyTorch library \cite{begaint2020compressai} is used as platform to implement the proposed method. The Flickr2W \cite{liu2020unified} is used as the training dataset. For simplicity, images smaller than 256 pixels in height or width were discarded. The remaining 20,718 images were then randomly cropped to 256$\times$256 patches and used as input to train the model.

The proposed method is trained in an end-to-end way. The objective function is the rate-distortion cost,
  \begin{equation}
    \begin{aligned}
        L &= R(\hat{y}) + \lambda \cdot D(x, \hat{x}) \\
          &= \mathbb{E}_{x \sim p{(x)}} \left[-\log_2 p(\hat{y}|\hat{z}) - \log_2 p(\hat{z}) \right] \\
          &\quad + \lambda \cdot \mathbb{E}_{x \sim p{(x)}} [d(x, \hat{x})]
    \end{aligned}
  \end{equation}
where the rate \( R \) represents the entropy of \(\hat{y}\), estimated through an entropy model \cite{minnen2018joint}, and the hyperprior \(\hat{z}\). \( D \) denotes the distortion, measured as Mean Square Error (MSE) between the input and reconstructed images. \( \lambda \) is the Lagrange multiplier to balance the rate and distortion. Different \( \lambda s \) are used including \( (0.0032, 0.0075, 0.015, 0.03, 0.045, 0.09) \) to achieve different bitrate points. Adam \cite{kingma2014adam} is used as the optimizer. The initial learning rate is set to \( 4 \times 10^{-4} \) and gradually reduced to \( 1 \times 10^{-6} \) based on the validation accuracy of a small percentage of training data (200 images are selected as the validation set). The batch size is set to 24.

\begin{figure*}[tbp]
  \centering
  \begin{subfigure}{0.245\textwidth} 
    \centering
    \adjustbox{valign=m}{\includegraphics[width=\linewidth]{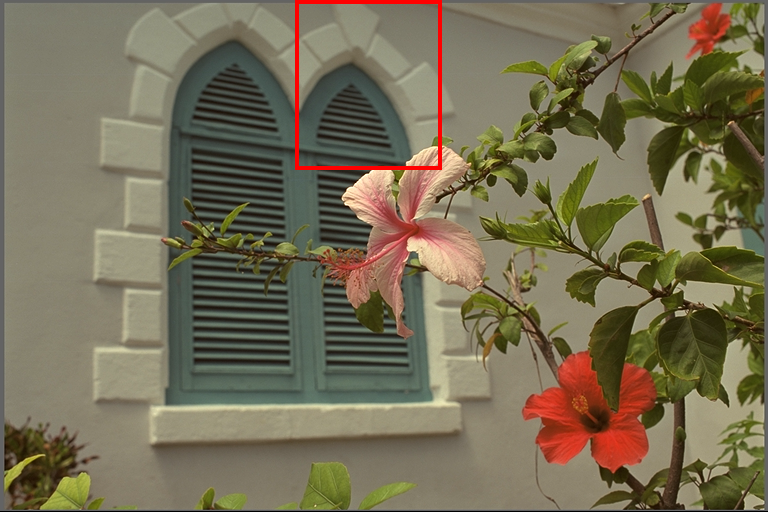}}
    \caption*{Original Image} 
  \end{subfigure}
  \hfill
  \begin{subfigure}{0.245\textwidth} 
    \centering
    \adjustbox{valign=m}{\includegraphics[width=\linewidth]{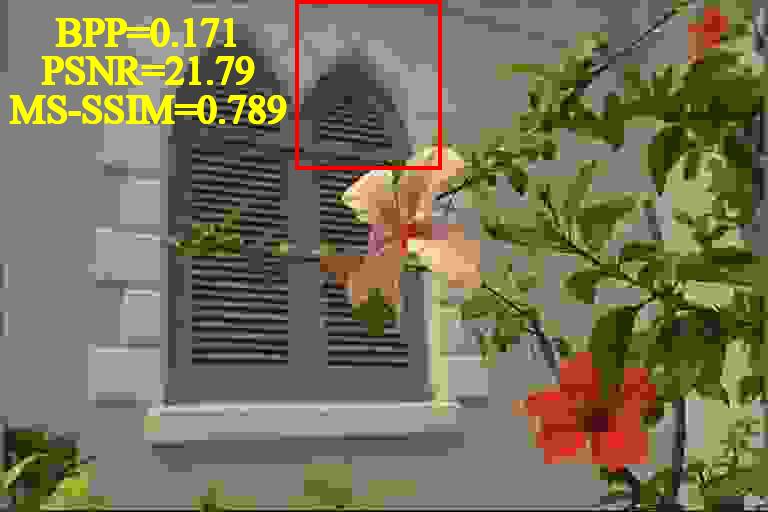}}
    \caption*{JPEG \cite{wallace1991jpeg}} 
  \end{subfigure}
  \hfill
  \begin{subfigure}{0.245\textwidth} 
    \centering
    \adjustbox{valign=m}{\includegraphics[width=\linewidth]{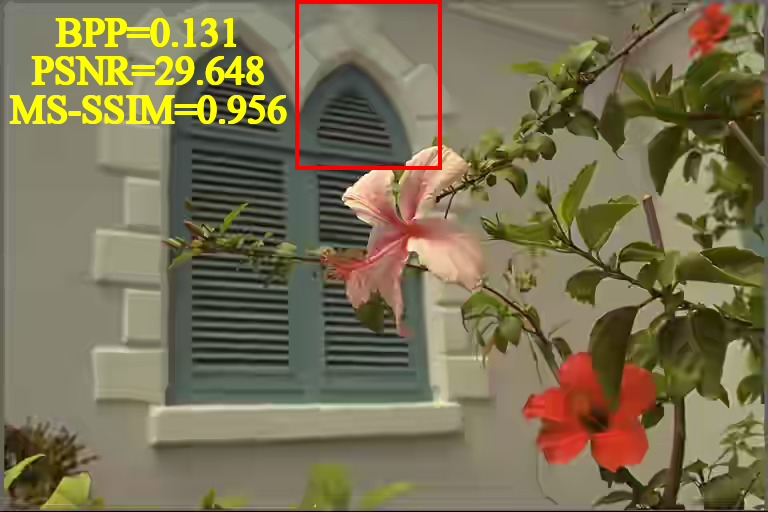}}
    \caption*{BPG \cite{bellard2015bpg}} 
  \end{subfigure}
  \hfill
  \begin{subfigure}{0.245\textwidth} 
    \centering
    \adjustbox{valign=m}{\includegraphics[width=\linewidth]{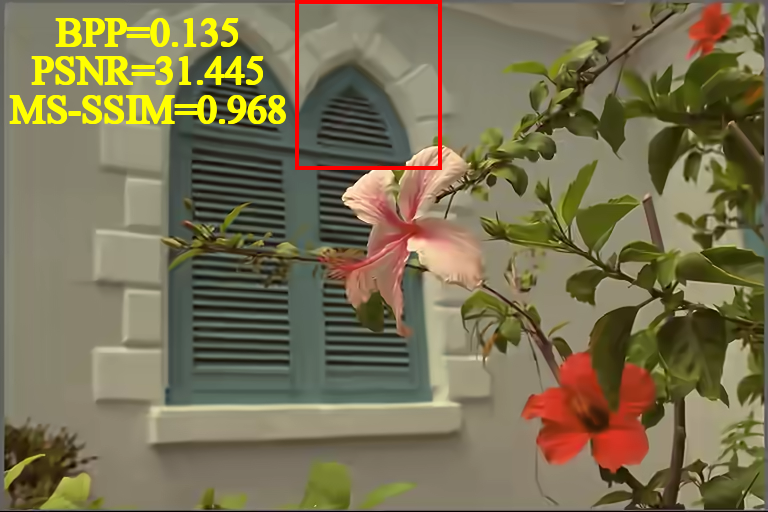}}
    \caption*{VTM12.1 \cite{bross2021overview}} 
  \end{subfigure}

  \vspace{0.1cm} 

  \begin{subfigure}{0.245\textwidth} 
    \centering
    \adjustbox{valign=m}{\includegraphics[width=\linewidth]{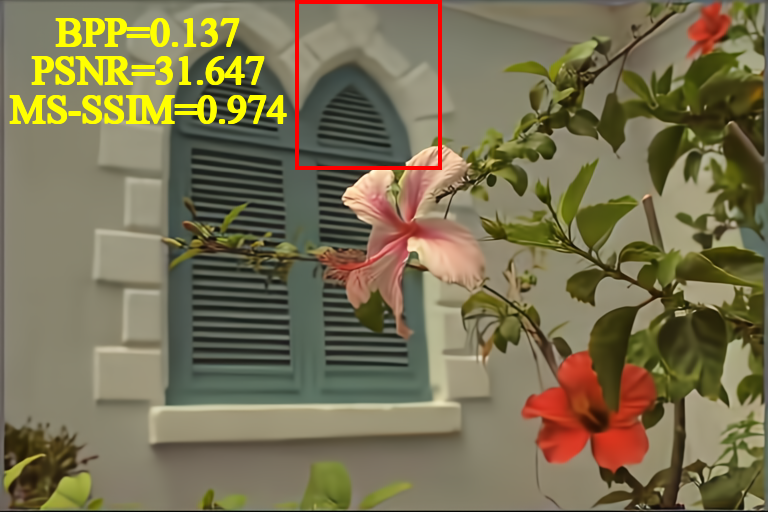}}
    \caption*{Cheng2020 \cite{cheng2020learned}} 
  \end{subfigure}
  \hfill
  \begin{subfigure}{0.245\textwidth} 
    \centering
    \adjustbox{valign=m}{\includegraphics[width=\linewidth]{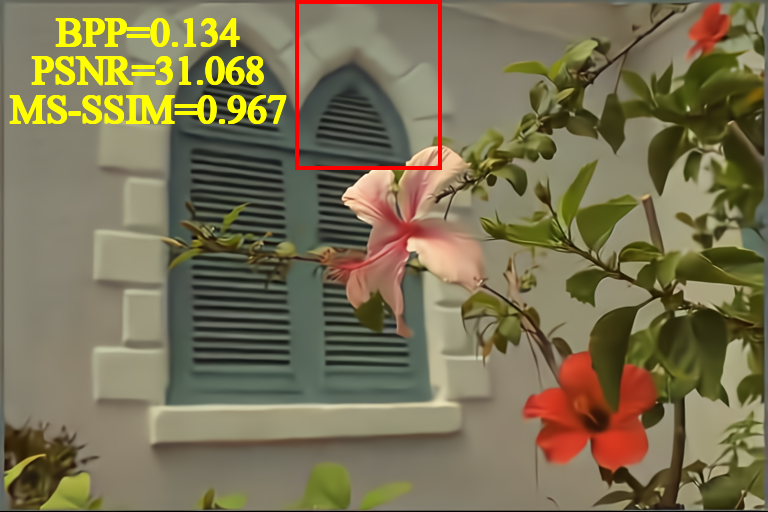}}
    \caption*{ANFIC \cite{ho2021anfic}} 
  \end{subfigure}
  \hfill
  \begin{subfigure}{0.245\textwidth} 
    \centering
    \adjustbox{valign=m}{\includegraphics[width=\linewidth]{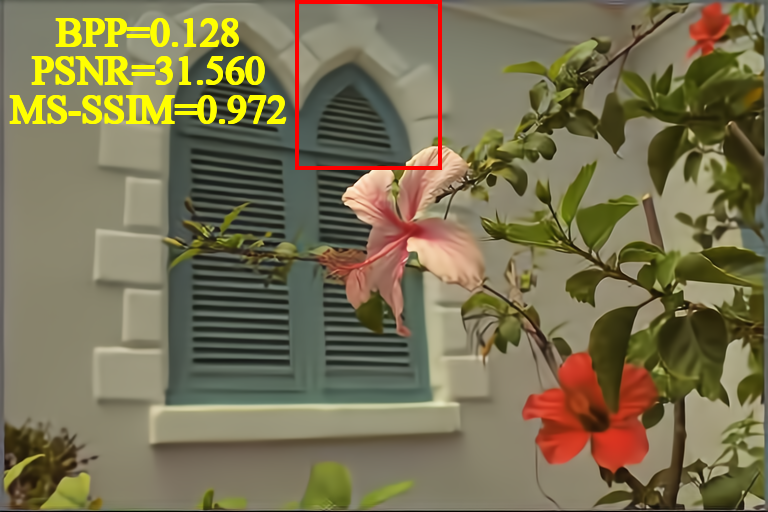}}
    \caption*{InvCompress \cite{xie2021enhanced}} 
  \end{subfigure}
  \hfill
  \begin{subfigure}{0.245\textwidth} 
    \centering
    \adjustbox{valign=m}{\includegraphics[width=\linewidth]{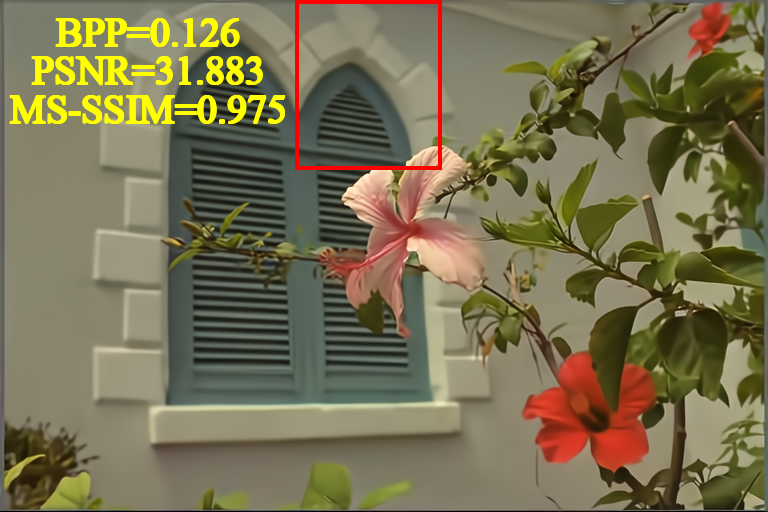}}
    \caption*{Ours} 
  \end{subfigure}  

  \vspace{0.1cm} 

  \begin{subfigure}{0.115\textwidth} 
    \centering
    \adjustbox{valign=m}{\includegraphics[width=\linewidth]{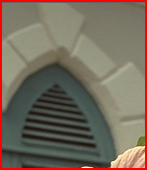}}
    \caption*{Original Image} 
  \end{subfigure}
  \hfill
  \begin{subfigure}{0.115\textwidth} 
    \centering
    \adjustbox{valign=m}{\includegraphics[width=\linewidth]{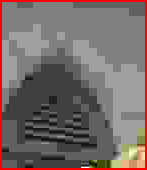}}
    \caption*{JPEG} 
  \end{subfigure}
  \hfill
  \begin{subfigure}{0.115\textwidth} 
    \centering
    \adjustbox{valign=m}{\includegraphics[width=\linewidth]{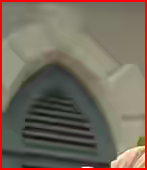}}
    \caption*{BPG} 
  \end{subfigure}
  \hfill
  \begin{subfigure}{0.115\textwidth} 
    \centering
    \adjustbox{valign=m}{\includegraphics[width=\linewidth]{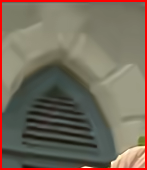}}
    \caption*{VTM12.1} 
  \end{subfigure}
  \hfill
  \begin{subfigure}{0.115\textwidth} 
    \centering
    \adjustbox{valign=m}{\includegraphics[width=\linewidth]{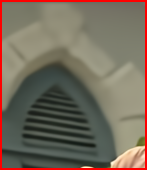}}
    \caption*{Cheng2020} 
  \end{subfigure}
  \hfill
  \begin{subfigure}{0.115\textwidth} 
    \centering
    \adjustbox{valign=m}{\includegraphics[width=\linewidth]{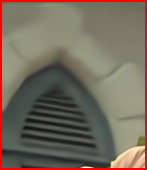}}
    \caption*{ANFIC} 
  \end{subfigure}
  \hfill
  \begin{subfigure}{0.115\textwidth} 
    \centering
    \adjustbox{valign=m}{\includegraphics[width=\linewidth]{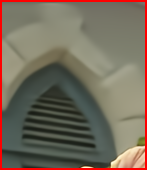}}
    \caption*{InvCompress} 
  \end{subfigure}
  \hfill
  \begin{subfigure}{0.115\textwidth} 
    \centering
    \adjustbox{valign=m}{\includegraphics[width=\linewidth]{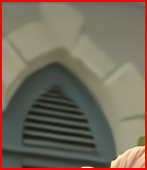}}
    \caption*{Ours} 
  \end{subfigure}

  \caption{Qualitative comparison of the proposed A-INN against the existing methods on the reconstructed image \textit{kodim07} from Kodak dataset.} 
  \label{fig:visual-kodim07}
\end{figure*}

\textbf{Testing details:} Four widely used benchmark datasets, including the Kodak dataset \cite{kodak1999}, CLIC 2020 Professional Test dataset, CLIC 2020 Mobile Test dataset (also shorten as CLIC) \cite{toderici2020workshop}, and Tecnick dataset \cite{asuni2014testimages}, are used to comprehensively evaluate the proposed method. These datasets contain images with varying resolutions and content types. The Kodak dataset consists of 24 images with resolution of \( 768 \times 512 \). The CLIC 2020 Professional Test dataset includes 250 images while the CLIC 2020 Mobile Test dataset comprises 178 images with resolutions of \( 2016 \times 1512 \). Additionally, the Tecnick dataset contains 100 images, each with a resolution of \( 1200 \times 1200 \). During the testing phase, the bits obtained by the entropy coding are used, in contrast to the training process where the bits are estimated with entropy. The distortion is measured in the same way using MSE.

\textbf{Evaluation Metrics:} BD-rate saving, based on the rate and quality (distortion), is used to evaluate the performance of our A-INN over existing methods. The rate is measured in bits per pixel (bpp), while the quality is measured in Peak Signal-to-Noise Ratio (PSNR), respectively.

\subsection{Rate-Distortion Performance Comparison with State-of-the-art Methods}

To evaluate the performance of our method, existing methods, including INN-based methods \cite{xie2021enhanced, ho2021anfic}, other state-of-the-art learned image compression methods \cite{cheng2020learned, zou2022devil}, and traditional compression standards \cite{wallace1991jpeg, christopoulos2000jpeg2000, google2010webp, bellard2015bpg, chen2018overview, sullivan2012overview, bross2021overview}, are used for comparison. Table \ref{tab:traditional-compression} and Table \ref{tab:learning-based-methods} present the result comparison of the proposed method over the traditional methods and learned image compression methods, respectively. The compared traditional methods include JPEG \cite{wallace1991jpeg} and its successor JPEG2000 \cite{christopoulos2000jpeg2000}, as well as emerging formats like WebP \cite{google2010webp}, BPG \cite{bellard2015bpg}, and AV1 \cite{chen2018overview}, HM \cite{sullivan2012overview}, and VVC (VTM 12.1) \cite{bross2021overview} codecs. In the comparison, BPG444 \cite{bellard2015bpg} is used as the baseline when calculating the BD-Rate saving. From Table \ref{tab:traditional-compression}, it is clear that the proposed A-INN method outperforms all traditional compression methods including the recent VTM12.1. The A-INN method further achieves an average BD-rate savings of 6.31\% over VTM12.1, which is a significant improvement in terms of compression efficiency.

\begin{table}[tbp]
  \centering
  \caption{Comparison of the proposed method against the traditional compression methods in terms of BD-Rate saving}
  \setlength{\tabcolsep}{11pt} 
  \begin{tabular}{lcccc}
  \hline
  \multirow{2}{*}{Methods} & \multicolumn{3}{c}{BD-Rate (\%)} \\ 
  \cline{2-4}
  & Kodak & CLIC & CLIC.P \\ 
  \hline
  JPEG \cite{wallace1991jpeg}      & 144.51       & 143.96 & 185.06 \\
  JPEG2000 \cite{christopoulos2000jpeg2000}  & 55.19        & 53.14  & 54.58  \\
  WebP \cite{google2010webp}      & 43.70        & 47.11  & 65.49  \\
  BPG444 \cite{bellard2015bpg}    & 0            & 0      & 0   \\
  AV1 \cite{chen2018overview}       & -6.26        & -8.91  & -13.00 \\
  HM \cite{sullivan2012overview}        & 0.81         & 1.02   & -2.16  \\
  VTM12.1 \cite{bross2021overview}   & -18.07       & -16.64 & -22.27 \\
  \hline
  \bfseries Ours          & \bfseries -24.39       & \bfseries -22.20 & \bfseries -29.32 \\ \hline
  \end{tabular}
  \label{tab:traditional-compression}
\end{table}

\begin{table}[tbp]
  \centering
  \caption{Comparison of  the proposed method against the existing learned image comparison methods in terms of BD-Rate saving}
  \begin{tabular}{l@{\hskip 10pt}c@{\hskip 10pt}c@{\hskip 10pt}c@{\hskip 10pt}c}
  \hline
  \multirow{2}{*}{Methods} & \multicolumn{4}{c}{BD-Rate (\%)} \\ 
  \cline{2-5}
  & Kodak & Tecnick & CLIC & CLIC.P \\
  \hline
  Ballé2017 \cite{balle2016end} & 60.18 & 74.48 & 91.42 & 74.41 \\
  Ballé2018 \cite{balle2018variational} & 30.87 & 30.90 & 40.11 & 34.02 \\
  Minnen2018 \cite{minnen2018joint} & 15.79 & 18.67 & 5.64 & 22.63 \\
  Cheng2020 \cite{cheng2020learned} & 0 & 0 & 0 & 0 \\
  ANFIC \cite{ho2021anfic} & -0.62 & -2.61 & 0.28 & 1.02 \\
  STF \cite{zou2022devil} & -8.76 & -10.06 & -7.56 & -9.60 \\
  InvCompress \cite{xie2021enhanced} & -5.96 & -5.90 & -8.65 & -8.95 \\
  \hline
  \textbf{Ours} & \textbf{-11.39} & \textbf{-10.74} & \textbf{-14.19} & \textbf{-14.30} \\
  \hline
  \end{tabular}
  \label{tab:learning-based-methods}
\end{table}

\begin{figure*}[tbp]
  \centering
  \begin{subfigure}{0.245\textwidth} 
    \centering
    \adjustbox{valign=m}{\includegraphics[width=\linewidth]{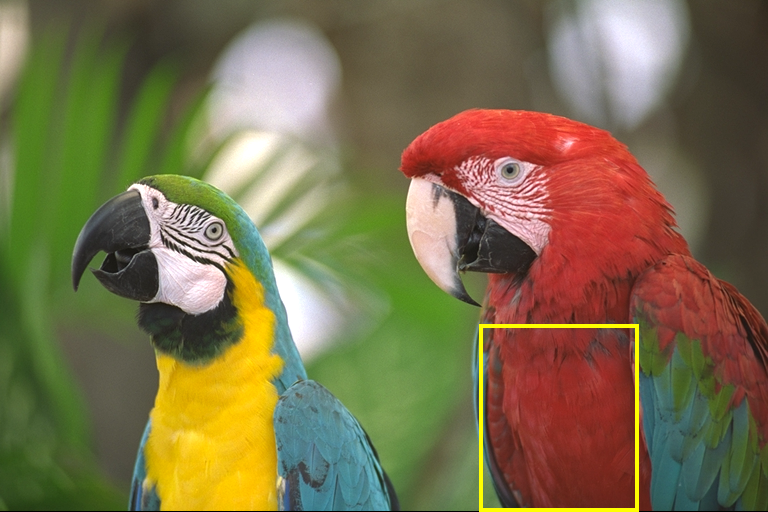}}
    \caption*{Original Image} 
  \end{subfigure}
  \hfill
  \begin{subfigure}{0.245\textwidth} 
    \centering
    \adjustbox{valign=m}{\includegraphics[width=\linewidth]{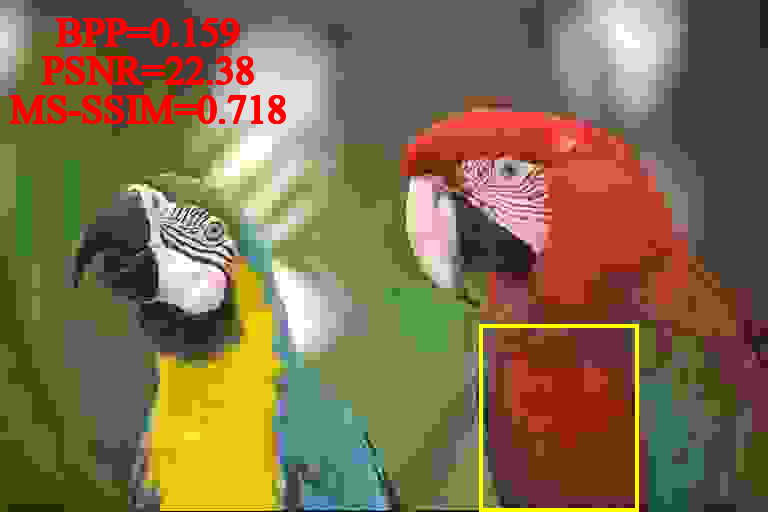}}
    \caption*{JPEG \cite{wallace1991jpeg}} 
  \end{subfigure}
  \hfill
  \begin{subfigure}{0.245\textwidth} 
    \centering
    \adjustbox{valign=m}{\includegraphics[width=\linewidth]{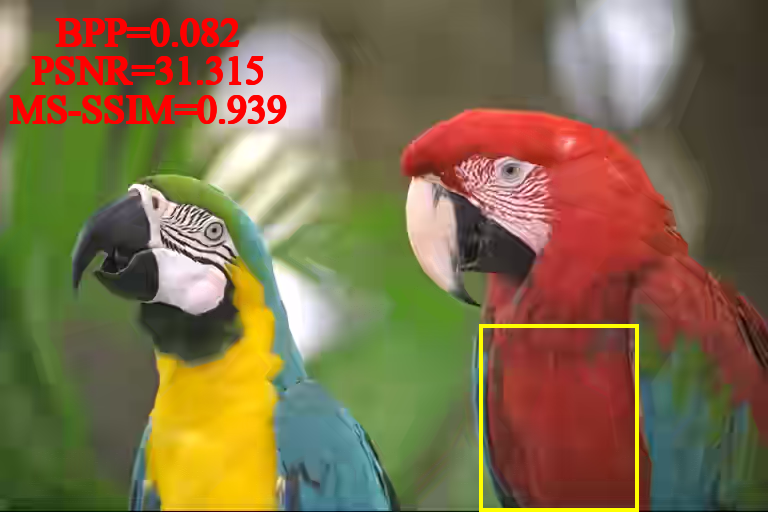}}
    \caption*{BPG \cite{bellard2015bpg}} 
  \end{subfigure}
  \hfill
  \begin{subfigure}{0.245\textwidth} 
    \centering
    \adjustbox{valign=m}{\includegraphics[width=\linewidth]{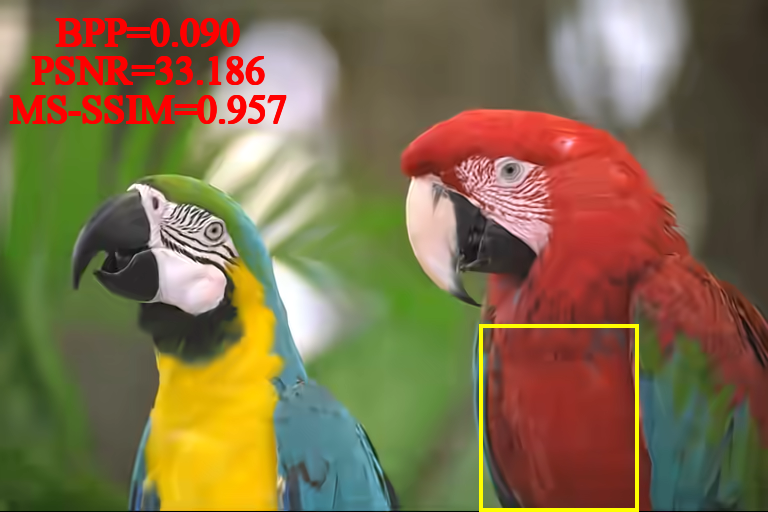}}
    \caption*{VTM12.1 \cite{bross2021overview}} 
  \end{subfigure}

  \vspace{0.1cm} 

  \begin{subfigure}{0.245\textwidth} 
    \centering
    \adjustbox{valign=m}{\includegraphics[width=\linewidth]{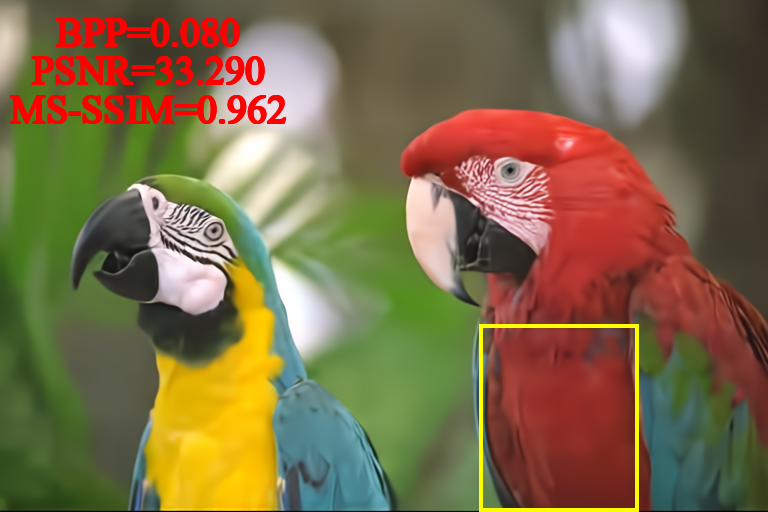}}
    \caption*{Cheng2020 \cite{cheng2020learned}} 
  \end{subfigure}
  \hfill
  \begin{subfigure}{0.245\textwidth} 
    \centering
    \adjustbox{valign=m}{\includegraphics[width=\linewidth]{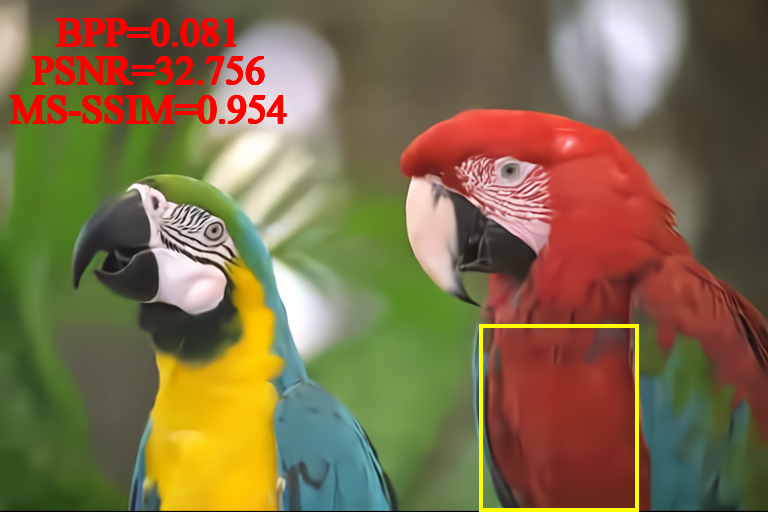}}
    \caption*{ANFIC \cite{ho2021anfic}} 
  \end{subfigure}
  \hfill
  \begin{subfigure}{0.245\textwidth} 
    \centering
    \adjustbox{valign=m}{\includegraphics[width=\linewidth]{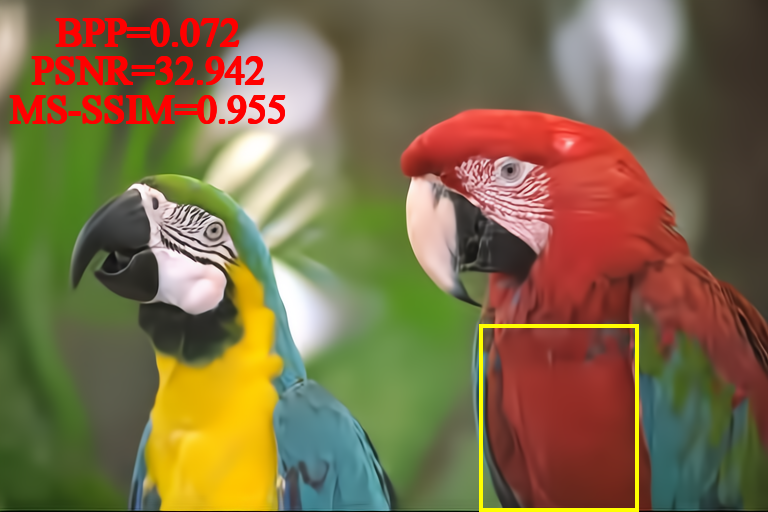}}
    \caption*{InvCompress \cite{xie2021enhanced}} 
  \end{subfigure}
  \hfill
  \begin{subfigure}{0.245\textwidth} 
    \centering
    \adjustbox{valign=m}{\includegraphics[width=\linewidth]{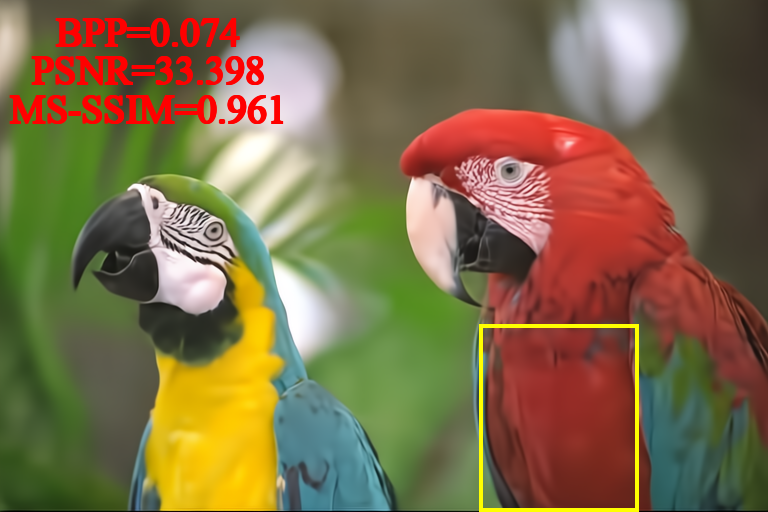}}
    \caption*{Ours} 
  \end{subfigure}  

  \vspace{0.1cm} 

  \begin{subfigure}{0.115\textwidth} 
    \centering
    \adjustbox{valign=m}{\includegraphics[width=\linewidth]{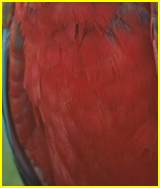}}
    \caption*{Original Image} 
  \end{subfigure}
  \hfill
  \begin{subfigure}{0.115\textwidth} 
    \centering
    \adjustbox{valign=m}{\includegraphics[width=\linewidth]{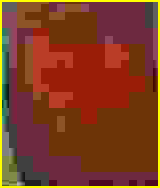}}
    \caption*{JPEG} 
  \end{subfigure}
  \hfill
  \begin{subfigure}{0.115\textwidth} 
    \centering
    \adjustbox{valign=m}{\includegraphics[width=\linewidth]{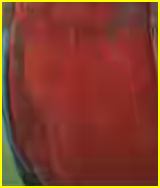}}
    \caption*{BPG} 
  \end{subfigure}
  \hfill
  \begin{subfigure}{0.115\textwidth} 
    \centering
    \adjustbox{valign=m}{\includegraphics[width=\linewidth]{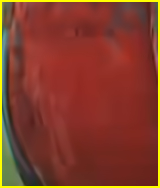}}
    \caption*{VTM12.1} 
  \end{subfigure}
  \hfill
  \begin{subfigure}{0.115\textwidth} 
    \centering
    \adjustbox{valign=m}{\includegraphics[width=\linewidth]{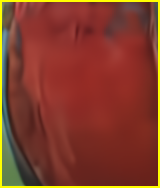}}
    \caption*{Cheng2020} 
  \end{subfigure}
  \hfill
  \begin{subfigure}{0.115\textwidth} 
    \centering
    \adjustbox{valign=m}{\includegraphics[width=\linewidth]{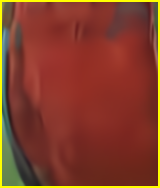}}
    \caption*{ANFIC} 
  \end{subfigure}
  \hfill
  \begin{subfigure}{0.115\textwidth} 
    \centering
    \adjustbox{valign=m}{\includegraphics[width=\linewidth]{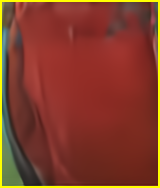}}
    \caption*{InvCompress} 
  \end{subfigure}
  \hfill
  \begin{subfigure}{0.115\textwidth} 
    \centering
    \adjustbox{valign=m}{\includegraphics[width=\linewidth]{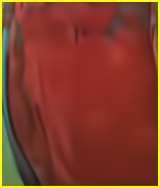}}
    \caption*{Ours} 
  \end{subfigure}

  \caption{Qualitative comparison of the proposed A-INN against the existing methods on the reconstructed image \textit{kodim23} from Kodak dataset.} 
  \label{fig:visual-kodim23}
\end{figure*}

For learned image compression models, methods including Ballé et al. \cite{balle2016end}, Ballé et al. \cite{balle2018variational}, Minnen et al. \cite{minnen2018joint}, Cheng et al. \cite{cheng2020learned}, Ho et al. \cite{ho2021anfic} (ANFIC), Zou et al. \cite{zou2022devil} (STF), and Xie et al. \cite{xie2021enhanced} (InvCompress) are compared. The relevant data points for this comparison were sourced from the CompressAI \cite{begaint2020compressai}, which provides a comprehensive benchmark for assessing the performance of our model within the current landscape of image compression techniques. In the comparison, Cheng et al. \cite{cheng2020learned} is used as the baseline to calculate the BD-Rate saving. From Table \ref{tab:learning-based-methods}, it can be seen that our A-INN model achieves the best results over the existing learned image compression methods, achieving an average BD-Rate saving of 12.66\%. Compared with the INN-based method \cite{xie2021enhanced}, our A-INN achieves an average BD-rate saving of 5.29\%, validating the effectiveness of the proposed method. Note that the proposed method uses the same coupling layer as \cite{xie2021enhanced} for a fair comparison, which is a basic architecture and the performance can be further improved with better and complicated architectures. The RD curve comparison over different methods are shown in Fig. \ref{fig:rd-curv}. It can be seen that the proposed method achieves the best result over the existing methods with clearly better RD performance.

Furthermore, the visual assessments comparing the reconstructed images using our method with those produced by existing methods at equivalent bit rates are also provided as shown in Figs. \ref{fig:visual-kodim07} and \ref{fig:visual-kodim23}. Considering the rate is implicitly controlled with the Lagrange multiplier, there exist small differences among the rates of different methods. For fair comparison, the rate of our method is smaller than the other methods. From Figs. \ref{fig:visual-kodim07} and \ref{fig:visual-kodim23}, it can be seen that the proposed method achieves better result than the existing methods in terms of visual fidelity and the retention of intricate details, effectively highlighting the strength of our proposed approach.

\begin{table}[tbp]
  \centering
  \caption{Ablation study on the different proposed modules in terms of BD-Rate saving}
  \begin{tabular}{p{4.53cm}@{\hskip 2.6pt}c@{\hskip 2.6pt}c@{\hskip 2.6pt}c@{\hskip 2.6pt}c@{\hskip 2.6pt}c}
  \hline
  Modules & \( M_a \) & \( M_b \) & \( M_c \) & \( M_d \) & \( M_e \) \\
  \hline
  Residual block-based Denoising & \(\checkmark\) & & & &  \\
  Dense block-based Denoising & & \(\checkmark\) & & &  \\
  Structure-based Denoising & & & \(\checkmark\) & & \\
  CFRM & & & & \(\checkmark\) & \\
  FDSM & & & & & \(\checkmark\) \\
  \hline
  BD-rate (\%) & -0.35 & -0.8 & -2.1 & -2.28 & -0.65 \\
  \hline
  \end{tabular}
  \label{tab:module_comparison}
\end{table}

\subsection{Ablation Studies}
In this subsection, ablation studies are conducted to examine the individual modules within the proposed A-INN. A-INN using the plain invertible blocks without the proposed modules are used as the baseline, and each module is added on top of the baseline to test its performance.

\vspace{0.2cm}
\textbf{Evaluation on the Progressive Denoising Module (PDM).} To evaluate the effect of using PDM in A-INN and also the difference among different types of denoising functions, experiments using different denoising modules are performed. The results are shown in Table \ref{tab:module_comparison}. It can be seen that all denoising modules, including the residual block-based, dense block-based and structure-based, can improve the performance. Moreover, the structure-based denoising module with down-sampling to extract the larger scale structure information clearly achieves better performance than the general residual block-based and dense block-based, achieving over 2\% BD-Rate saving than the baseline.

\vspace{0.2cm}
\textbf{Evaluation on the Cascaded Feature Recovery Module (CFRM).} To evaluate the proposed CFRM in reducing the noise introduced in feature channel squeezing, ablation experiment on CFRM is conducted and the results are shown in Table \ref{tab:module_comparison}. It can be seen that the result of using CFRM achieves a BD-Rate saving of 2.28\%, validating its effectiveness. This also verifies the necessity of a denoising module corresponding to the noise introduced in channel squeezing.

\begin{figure}[tbp]
	\centering
	\includegraphics[width=0.5\textwidth]{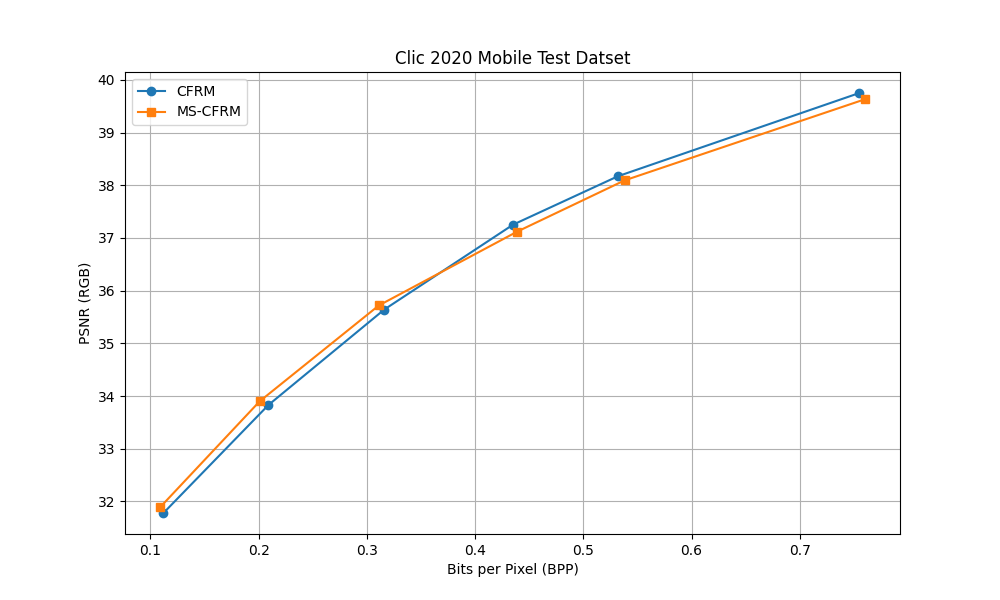}
  \abovecaptionskip=0pt 
  \belowcaptionskip=0pt 
	\caption{RD curve comparison between CFRM and MS-CFRM.}
  \label{fig:ablation}
\end{figure}

Moreover, Multi-Stage Cascaded Feature Recovery Module (MS-CFRM) is further evaluated, which gradually reduces and restores the feature channels at the encoder and decoder, respectively, in order to reduce the complexity at the later stage of encoding. For INN-based coding, the number of feature channels are increased along with the decrease in the spatial dimension. The number of channels grows to 768 ($3 \times 4^4$) with 4 down-shuffle operations. The same setting as in \cite{xie2021enhanced} is used, where the channel number is set to 128 at lower-rate models (corresponding to the first three Lagrange multipliers) while set to 192 at higher-rate models (corresponding to the remaining last three Lagrange multipliers), resulting in $6\times$ and $4\times$ channel squeezing before entropy coding, respectively. This results in the oversmoothing problem and complex feature processing at the last stage. For MS-CFRM, multi-stage channel squeezing and corresponding channel up-sampling are used (two stages in the experiments) to reduce the above problem. The RD curve of the MS-CFRM against CFRM is shown in Fig. \ref{fig:ablation}. It can be seen that at lower bitrates with large channel reduction ratio ($6\times$), the MS-CFRM performs better than CFRM, while at higher bitrates with a relatively small channel reduction ratio, CFRM performs slightly better than MS-CFRM. Therefore, in the experiments, for higher-rate models with 192 channels of the latent feature, CFRM is used while MS-CFRM is used for the remaining lower-rate models.

\textbf{Evaluation on the Frequency Decomposition and Synthesis Module (FDSM).} The proposed FDSM is also evaluated individually, to verify the effectiveness of using frequency-enhanced representation in A-INN. The results are also shown in Table \ref{tab:module_comparison}. It can be seen that the proposed FDSM also improves the coding performance by 0.65\% in terms of BD-Rate saving.

\section{Conclusion}
This paper presents the Approximately Invertible Neural Network (A-INN), a novel framework for learned image compression. A-INN is formulated for rate-distortion optimization considering quantization in lossy image compression, which is different from using INN for image generative modeling. Three components, including a Progressive Denoising Module (PDM), a Cascaded Feature Recovery Module (CFRM) and a Frequency Decomposition and Synthesis Module (FDSM), are developed. Among them, the PDM and CFRM in A-INN are used to reduce the quantization noise and feature channel compression loss to further enhance the coupling between the invertible encoder and decoder. FDSM is designed with a dual-frequency synthesis to reduce the high-frequency information loss. Ablation studies have verified the effectiveness of each proposed module in A-INN. Experimental results demonstrate that the proposed A-INN outperforms the existing learned image compression methods, with 5.29\% BD-rate savings in average over the baseline.

{
  \bibliographystyle{ieeetr} 
  \bibliography{references} 

\begin{thebibliography}{10}

\bibitem{MaZJZWW20}
S.~Ma, X.~Zhang, C.~Jia, Z.~Zhao, S.~Wang, and S.~Wang, ``Image and video
  compression with neural networks: {A} review,'' {\em {IEEE} Trans. Circuits
  Syst. Video Technol.}, vol.~30, no.~6, pp.~1683--1698, 2020.

\bibitem{10478821}
X.~Liu, M.~Wang, S.~Wang, and S.~Kwong, ``Bilateral context modeling for
  residual coding in lossless 3d medical image compression,'' {\em IEEE
  Transactions on Image Processing}, vol.~33, pp.~2502--2513, 2024.

\bibitem{9694511}
Z.~Chen, S.~Gu, G.~Lu, and D.~Xu, ``Exploiting intra-slice and inter-slice
  redundancy for learning-based lossless volumetric image compression,'' {\em
  IEEE Transactions on Image Processing}, vol.~31, pp.~1697--1707, 2022.

\bibitem{LeiL0J0G22}
J.~Lei, X.~Liu, B.~Peng, D.~Jin, W.~Li, and J.~Gu, ``Deep stereo image
  compression via bi-directional coding,'' in {\em {IEEE/CVF} Conference on
  Computer Vision and Pattern Recognition, {CVPR} 2022, New Orleans, LA, USA,
  June 18-24, 2022}, pp.~19637--19646, {IEEE}, 2022.

\bibitem{9646488}
T.~Dardouri, M.~Kaaniche, A.~Benazza-Benyahia, and J.-C. Pesquet, ``Dynamic
  neural network for lossy-to-lossless image coding,'' {\em IEEE Transactions
  on Image Processing}, vol.~31, pp.~569--584, 2022.

\bibitem{10121344}
F.~Kamisli, ``Learned lossless image compression through interpolation with low
  complexity,'' {\em IEEE Transactions on Circuits and Systems for Video
  Technology}, vol.~33, no.~12, pp.~7832--7841, 2023.

\bibitem{wallace1991jpeg}
G.~K. Wallace, ``The jpeg still picture compression standard,'' {\em
  Communications of the ACM}, vol.~34, no.~4, pp.~30--44, 1991.

\bibitem{bross2021overview}
B.~Bross, Y.-K. Wang, Y.~Ye, S.~Liu, J.~Chen, G.~J. Sullivan, and J.-R. Ohm,
  ``Overview of the versatile video coding (vvc) standard and its
  applications,'' {\em IEEE Transactions on Circuits and Systems for Video
  Technology}, vol.~31, no.~10, pp.~3736--3764, 2021.

\bibitem{9205606}
X.~Zhang, C.~Yang, X.~Li, S.~Liu, H.~Yang, I.~Katsavounidis, S.-M. Lei, and
  C.-C.~J. Kuo, ``Image coding with data-driven transforms: Methodology,
  performance and potential,'' {\em IEEE Transactions on Image Processing},
  vol.~29, pp.~9292--9304, 2020.

\bibitem{9359473}
T.~Chen, H.~Liu, Z.~Ma, Q.~Shen, X.~Cao, and Y.~Wang, ``End-to-end learnt image
  compression via non-local attention optimization and improved context
  modeling,'' {\em IEEE Transactions on Image Processing}, vol.~30,
  pp.~3179--3191, 2021.

\bibitem{DuanLMHMZ24}
Z.~Duan, M.~Lu, J.~Ma, Y.~Huang, Z.~Ma, and F.~Zhu, ``{QARV:}
  quantization-aware resnet {VAE} for lossy image compression,'' {\em {IEEE}
  Trans. Pattern Anal. Mach. Intell.}, vol.~46, no.~1, pp.~436--450, 2024.

\bibitem{ChenM23}
T.~Chen and Z.~Ma, ``Toward robust neural image compression: Adversarial attack
  and model finetuning,'' {\em {IEEE} Trans. Circuits Syst. Video Technol.},
  vol.~33, no.~12, pp.~7842--7856, 2023.

\bibitem{9455349}
Z.~Guo, Z.~Zhang, R.~Feng, and Z.~Chen, ``Causal contextual prediction for
  learned image compression,'' {\em IEEE Transactions on Circuits and Systems
  for Video Technology}, vol.~32, no.~4, pp.~2329--2341, 2022.

\bibitem{9568930}
Y.~Wu, X.~Li, Z.~Zhang, X.~Jin, and Z.~Chen, ``Learned block-based hybrid image
  compression,'' {\em IEEE Transactions on Circuits and Systems for Video
  Technology}, vol.~32, no.~6, pp.~3978--3990, 2022.

\bibitem{christopoulos2000jpeg2000}
C.~A. Christopoulos, T.~Ebrahimi, and A.~N. Skodras, ``Jpeg2000: the new still
  picture compression standard,'' in {\em Proceedings of the 2000 ACM workshops
  on Multimedia}, pp.~45--49, 2000.

\bibitem{sullivan2012overview}
G.~J. Sullivan, J.-R. Ohm, W.-J. Han, and T.~Wiegand, ``Overview of the high
  efficiency video coding (hevc) standard,'' {\em IEEE Transactions on circuits
  and systems for video technology}, vol.~22, no.~12, pp.~1649--1668, 2012.

\bibitem{xie2021enhanced}
Y.~Xie, K.~L. Cheng, and Q.~Chen, ``Enhanced invertible encoding for learned
  image compression,'' in {\em Proceedings of the 29th ACM international
  conference on multimedia}, pp.~162--170, 2021.

\bibitem{JinLPLLH22}
D.~Jin, J.~Lei, B.~Peng, W.~Li, N.~Ling, and Q.~Huang, ``Deep affine motion
  compensation network for inter prediction in {VVC},'' {\em {IEEE} Trans.
  Circuits Syst. Video Technol.}, vol.~32, no.~6, pp.~3923--3933, 2022.

\bibitem{JinLPPLL23}
D.~Jin, J.~Lei, B.~Peng, Z.~Pan, L.~Li, and N.~Ling, ``Learned video
  compression with efficient temporal context learning,'' {\em {IEEE} Trans.
  Image Process.}, vol.~32, pp.~3188--3198, 2023.

\bibitem{9585549}
N.~Yan, C.~Gao, D.~Liu, H.~Li, L.~Li, and F.~Wu, ``Sssic: Semantics-to-signal
  scalable image coding with learned structural representations,'' {\em IEEE
  Transactions on Image Processing}, vol.~30, pp.~8939--8954, 2021.

\bibitem{9934922}
M.~Li, L.~Shen, Y.~Lin, K.~Wang, and J.~Chen, ``Extreme underwater image
  compression using physical priors,'' {\em IEEE Transactions on Circuits and
  Systems for Video Technology}, vol.~33, no.~4, pp.~1937--1951, 2023.

\bibitem{9984191}
Z.~Fang, L.~Shen, M.~Li, Z.~Wang, and Y.~Jin, ``Prior-guided contrastive image
  compression for underwater machine vision,'' {\em IEEE Transactions on
  Circuits and Systems for Video Technology}, vol.~33, no.~6, pp.~2950--2961,
  2023.

\bibitem{9043584}
Z.~Jin, M.~Z. Iqbal, W.~Zou, X.~Li, and E.~Steinbach, ``Dual-stream multi-path
  recursive residual network for jpeg image compression artifacts reduction,''
  {\em IEEE Transactions on Circuits and Systems for Video Technology},
  vol.~31, no.~2, pp.~467--479, 2021.

\bibitem{9998500}
Y.~Shi, K.~Zhang, J.~Wang, N.~Ling, and B.~Yin, ``Variable-rate image
  compression based on side information compensation and r-$\lambda$ model rate
  control,'' {\em IEEE Transactions on Circuits and Systems for Video
  Technology}, vol.~33, no.~7, pp.~3488--3501, 2023.

\bibitem{10226331}
D.~Zhang, F.~Li, M.~Liu, R.~Cong, H.~Bai, M.~Wang, and Y.~Zhao, ``Exploring
  resolution fields for scalable image compression with uncertainty guidance,''
  {\em IEEE Transactions on Circuits and Systems for Video Technology},
  vol.~34, no.~4, pp.~2934--2948, 2024.

\bibitem{zou2022devil}
R.~Zou, C.~Song, and Z.~Zhang, ``The devil is in the details: Window-based
  attention for image compression,'' in {\em Proceedings of the IEEE/CVF
  conference on computer vision and pattern recognition}, pp.~17492--17501,
  2022.

\bibitem{9858899}
Z.~Tang, H.~Wang, X.~Yi, Y.~Zhang, S.~Kwong, and C.-C.~J. Kuo, ``Joint graph
  attention and asymmetric convolutional neural network for deep image
  compression,'' {\em IEEE Transactions on Circuits and Systems for Video
  Technology}, vol.~33, no.~1, pp.~421--433, 2023.

\bibitem{ZhangZT23}
G.~Zhang, X.~Zhang, and L.~Tang, ``Enhanced quantified local implicit neural
  representation for image compression,'' {\em {IEEE} Signal Process. Lett.},
  vol.~30, pp.~1742--1746, 2023.

\bibitem{kingma2013auto}
D.~P. Kingma and M.~Welling, ``Auto-encoding variational bayes,'' in {\em
  International Conference on Learning Representations, {ICLR} 2014}.

\bibitem{9989403}
S.~Li, H.~Li, W.~Dai, C.~Li, J.~Zou, and H.~Xiong, ``Learned progressive image
  compression with dead-zone quantizers,'' {\em IEEE Transactions on Circuits
  and Systems for Video Technology}, vol.~33, no.~6, pp.~2962--2978, 2023.

\bibitem{9144534}
D.~Mishra, S.~K. Singh, and R.~K. Singh, ``Wavelet-based deep auto
  encoder-decoder (wdaed)-based image compression,'' {\em IEEE Transactions on
  Circuits and Systems for Video Technology}, vol.~31, no.~4, pp.~1452--1462,
  2021.

\bibitem{10268865}
W.~Duan, Z.~Chang, C.~Jia, S.~Wang, S.~Ma, L.~Song, and W.~Gao, ``Learned image
  compression using cross-component attention mechanism,'' {\em IEEE
  Transactions on Image Processing}, vol.~32, pp.~5478--5493, 2023.

\bibitem{helminger2020lossy}
L.~Helminger, A.~Djelouah, M.~Gross, and C.~Schroers, ``Lossy image compression
  with normalizing flows,'' {\em arXiv preprint arXiv:2008.10486}, 2020.

\bibitem{google2010webp}
Google, ``Web picture format.''
  \url{https://chromium.googlesource.com/webm/libwebp}, 2010.

\bibitem{bellard2015bpg}
F.~Bellard, ``Bpg image format.'' \url{https://bellard.org/bpg/}, 2015.

\bibitem{chen2018overview}
Y.~Chen, D.~Murherjee, J.~Han, A.~Grange, Y.~Xu, Z.~Liu, S.~Parker, C.~Chen,
  H.~Su, U.~Joshi, {\em et~al.}, ``An overview of core coding tools in the av1
  video codec,'' in {\em 2018 picture coding symposium (PCS)}, pp.~41--45,
  IEEE, 2018.

\bibitem{ahmed1974discrete}
N.~Ahmed, T.~Natarajan, and K.~R. Rao, ``Discrete cosine transform,'' {\em IEEE
  transactions on Computers}, vol.~100, no.~1, pp.~90--93, 1974.

\bibitem{marpe2003context}
D.~Marpe, H.~Schwarz, and T.~Wiegand, ``Context-based adaptive binary
  arithmetic coding in the h. 264/avc video compression standard,'' {\em IEEE
  Transactions on circuits and systems for video technology}, vol.~13, no.~7,
  pp.~620--636, 2003.

\bibitem{netravali1980picture}
A.~N. Netravali and J.~O. Limb, ``Picture coding: A review,'' {\em Proceedings
  of the IEEE}, vol.~68, no.~3, pp.~366--406, 1980.

\bibitem{balle2016end}
J.~Ball{\'{e}}, V.~Laparra, and E.~P. Simoncelli, ``End-to-end optimized image
  compression,'' in {\em International Conference on Learning Representations,
  {ICLR} 2017}.

\bibitem{balle2018variational}
J.~Ball{\'{e}}, D.~Minnen, S.~Singh, S.~J. Hwang, and N.~Johnston,
  ``Variational image compression with a scale hyperprior,'' in {\em
  International Conference on Learning Representations, {ICLR} 2018}.

\bibitem{minnen2018joint}
D.~Minnen, J.~Ball{\'e}, and G.~D. Toderici, ``Joint autoregressive and
  hierarchical priors for learned image compression,'' {\em Advances in neural
  information processing systems}, vol.~31, 2018.

\bibitem{van2016conditional}
A.~Van~den Oord, N.~Kalchbrenner, L.~Espeholt, O.~Vinyals, A.~Graves, {\em
  et~al.}, ``Conditional image generation with pixelcnn decoders,'' {\em
  Advances in neural information processing systems}, vol.~29, 2016.

\bibitem{cheng2020learned}
Z.~Cheng, H.~Sun, M.~Takeuchi, and J.~Katto, ``Learned image compression with
  discretized gaussian mixture likelihoods and attention modules,'' in {\em
  Proceedings of the IEEE/CVF conference on computer vision and pattern
  recognition}, pp.~7939--7948, 2020.

\bibitem{lee2018context}
J.~Lee, S.~Cho, and S.~Beack, ``Context-adaptive entropy model for end-to-end
  optimized image compression,'' in {\em 7th International Conference on
  Learning Representations, {ICLR} 2019}.

\bibitem{zhou2019multi}
J.~Zhou, S.~Wen, A.~Nakagawa, K.~Kazui, and Z.~Tan, ``Multi-scale and
  context-adaptive entropy model for image compression,'' in {\em Proceedings
  of the IEEE/CVF conference on computer vision and pattern recognition}, 2019.

\bibitem{minnen2020channel}
D.~Minnen and S.~Singh, ``Channel-wise autoregressive entropy models for
  learned image compression,'' in {\em 2020 IEEE International Conference on
  Image Processing (ICIP)}, pp.~3339--3343, IEEE, 2020.

\bibitem{he2021checkerboard}
D.~He, Y.~Zheng, B.~Sun, Y.~Wang, and H.~Qin, ``Checkerboard context model for
  efficient learned image compression,'' in {\em Proceedings of the IEEE/CVF
  Conference on Computer Vision and Pattern Recognition}, pp.~14771--14780,
  2021.

\bibitem{li2020deep}
W.~Li, W.~Sun, Y.~Zhao, Z.~Yuan, and Y.~Liu, ``Deep image compression with
  residual learning,'' {\em Applied Sciences}, vol.~10, no.~11, p.~4023, 2020.

\bibitem{toderici2017full}
G.~Toderici, D.~Vincent, N.~Johnston, S.~Jin~Hwang, D.~Minnen, J.~Shor, and
  M.~Covell, ``Full resolution image compression with recurrent neural
  networks,'' in {\em Proceedings of the IEEE conference on Computer Vision and
  Pattern Recognition}, pp.~5306--5314, 2017.

\bibitem{johnston2018improved}
N.~Johnston, D.~Vincent, D.~Minnen, M.~Covell, S.~Singh, T.~Chinen, S.~J.
  Hwang, J.~Shor, and G.~Toderici, ``Improved lossy image compression with
  priming and spatially adaptive bit rates for recurrent networks,'' in {\em
  Proceedings of the IEEE conference on computer vision and pattern
  recognition}, pp.~4385--4393, 2018.

\bibitem{rippel2017real}
O.~Rippel and L.~Bourdev, ``Real-time adaptive image compression,'' in {\em
  International Conference on Machine Learning}, pp.~2922--2930, PMLR, 2017.

\bibitem{zhu2021transformer}
Y.~Zhu, Y.~Yang, and T.~Cohen, ``Transformer-based transform coding,'' in {\em
  International Conference on Learning Representations}, 2021.

\bibitem{dinh2014nice}
L.~Dinh, D.~Krueger, and Y.~Bengio, ``{NICE:} non-linear independent components
  estimation,'' in {\em International Conference on Learning Representations,
  {ICLR} 2015}.

\bibitem{dinh2016density}
L.~Dinh, J.~Sohl{-}Dickstein, and S.~Bengio, ``Density estimation using real
  {NVP},'' in {\em 5th International Conference on Learning Representations,
  {ICLR} 2017}.

\bibitem{kingma2018glow}
D.~P. Kingma and P.~Dhariwal, ``Glow: Generative flow with invertible 1x1
  convolutions,'' {\em Advances in neural information processing systems},
  vol.~31, 2018.

\bibitem{liang2021hierarchical}
J.~Liang, A.~Lugmayr, K.~Zhang, M.~Danelljan, L.~Van~Gool, and R.~Timofte,
  ``Hierarchical conditional flow: A unified framework for image
  super-resolution and image rescaling,'' in {\em Proceedings of the IEEE/CVF
  International Conference on Computer Vision}, pp.~4076--4085, 2021.

\bibitem{zhu2022high}
Y.~Zhu, C.~Wang, C.~Dong, K.~Zhang, H.~Gao, and C.~Yuan, ``High-frequency
  normalizing flow for image rescaling,'' {\em IEEE Transactions on Image
  Processing}, 2022.

\bibitem{wang2022low}
Y.~Wang, R.~Wan, W.~Yang, H.~Li, L.-P. Chau, and A.~Kot, ``Low-light image
  enhancement with normalizing flow,'' in {\em Proceedings of the AAAI
  conference on artificial intelligence}, vol.~36, pp.~2604--2612, 2022.

\bibitem{liu2021invertible}
Y.~Liu, Z.~Qin, S.~Anwar, P.~Ji, D.~Kim, S.~Caldwell, and T.~Gedeon,
  ``Invertible denoising network: A light solution for real noise removal,'' in
  {\em Proceedings of the IEEE/CVF conference on computer vision and pattern
  recognition}, pp.~13365--13374, 2021.

\bibitem{li2021dehazeflow}
H.~Li, J.~Li, D.~Zhao, and L.~Xu, ``Dehazeflow: Multi-scale conditional flow
  network for single image dehazing,'' in {\em Proceedings of the 29th ACM
  International Conference on Multimedia}, pp.~2577--2585, 2021.

\bibitem{wang2020modeling}
Y.~Wang, M.~Xiao, C.~Liu, S.~Zheng, and T.-Y. Liu, ``Modeling lost information
  in lossy image compression,'' {\em arXiv preprint arXiv:2006.11999}, 2020.

\bibitem{ho2021anfic}
Y.-H. Ho, C.-C. Chan, W.-H. Peng, H.-M. Hang, and M.~Doma{\'n}ski, ``Anfic:
  Image compression using augmented normalizing flows,'' {\em IEEE Open Journal
  of Circuits and Systems}, vol.~2, pp.~613--626, 2021.

\bibitem{huang2020augmented}
C.-W. Huang, L.~Dinh, and A.~Courville, ``Augmented normalizing flows: Bridging
  the gap between generative flows and latent variable models,'' {\em arXiv
  preprint arXiv:2002.07101}, 2020.

\bibitem{he2016deep}
K.~He, X.~Zhang, S.~Ren, and J.~Sun, ``Deep residual learning for image
  recognition,'' in {\em Proceedings of the IEEE conference on computer vision
  and pattern recognition}, pp.~770--778, 2016.

\bibitem{huang2017densely}
G.~Huang, Z.~Liu, L.~Van Der~Maaten, and K.~Q. Weinberger, ``Densely connected
  convolutional networks,'' in {\em Proceedings of the IEEE conference on
  computer vision and pattern recognition}, pp.~4700--4708, 2017.

\bibitem{ronneberger2015u}
O.~Ronneberger, P.~Fischer, and T.~Brox, ``U-net: Convolutional networks for
  biomedical image segmentation,'' in {\em Medical image computing and
  computer-assisted intervention--MICCAI 2015: 18th international conference,
  Munich, Germany, October 5-9, 2015, proceedings, part III 18}, pp.~234--241,
  Springer, 2015.

\bibitem{begaint2020compressai}
J.~B{\'e}gaint, F.~Racap{\'e}, S.~Feltman, and A.~Pushparaja, ``Compressai: a
  pytorch library and evaluation platform for end-to-end compression
  research,'' {\em arXiv preprint arXiv:2011.03029}, 2020.

\bibitem{liu2020unified}
J.~Liu, G.~Lu, Z.~Hu, and D.~Xu, ``A unified end-to-end framework for efficient
  deep image compression,'' {\em arXiv preprint arXiv:2002.03370}, 2020.

\bibitem{kingma2014adam}
D.~P. Kingma and J.~Ba, ``Adam: {A} method for stochastic optimization,'' in
  {\em 3rd International Conference on Learning Representations, {ICLR} 2015}.

\bibitem{kodak1999}
E.~K. Company, ``Kodak lossless true color image suite.''
  \url{http://r0k.us/graphics/kodak/}, 1999.

\bibitem{toderici2020workshop}
G.~Toderici, W.~Shi, R.~Timofte, L.~Theis, J.~Balle, E.~Agustsson, N.~Johnston,
  and F.~Mentzer, ``Workshop and challenge on learned image compression
  (clic2020),'' in {\em CVPR}, 2020.

\bibitem{asuni2014testimages}
N.~Asuni and A.~Giachetti, ``Testimages: a large-scale archive for testing
  visual devices and basic image processing algorithms.,'' in {\em STAG},
  pp.~63--70, 2014.

\end{thebibliography}
}

\end{document}